\documentclass[preprint]{aastex6}
\usepackage{amsmath,amssymb,amstext}
\usepackage{units}
\usepackage[all]{hypcap} 
\usepackage{natbib}
\usepackage{xspace}
\usepackage{graphicx}
\usepackage{todonotes}
\graphicspath{{./figures/}}

\shorttitle{Bolometric Lightcurves of SNe}
\shortauthors{Lusk \& Baron}

\begin{document}

\title{Bolometric Lightcurves of Peculiar Type II-P Supernovae}
\author{Jeremy A. Lusk\altaffilmark{1} and
  E.~Baron\altaffilmark{2,3}
} 
\affil{Homer L. Dodge Department of Physics and Astronomy, University of Oklahoma, Norman, OK 73019}

\altaffiltext{1}{Math/Natural Science Division, Midland College, Midland, TX 79705}
\altaffiltext{2}{Hamburger Sternwarte, Gojenbergsweg 112, 21029
  Hamburg, Germany}
\altaffiltext{3}
{
    Computational Cosmology Center, Computational Research Division, Lawrence Berkeley National Laboratory, 
    1 Cyclotron Road MS 50B-4206, Berkeley, CA, 94720
}

\begin{abstract}
  We examine the bolometric lightcurves of five Type~II-P supernovae (SNe 1998A, 2000cb, 2006V, 2006au and 2009E) which are thought to originate from blue supergiant progenitors using a new python package named \texttt{SuperBoL}.
  With this code, we calculate SNe lightcurves using three different techniques common in the literature: the \emph{quasi-bolometric} method, which integrates the observed photometry, the \emph{direct integration} method, which additionally corrects for unobserved flux in the UV and IR, and the \emph{bolometric correction} method, which uses correlations between observed colors and V-band bolometric corrections.
  We present here the lightcurves calculated by \texttt{SuperBoL} along with previously published lightcurves, as well as peak luminosities and $^{56}$Ni yields.
  We find that the direct integration and bolometric correction
  lightcurves largely agree with previously published lightcurves, but
  with what we believe to be more robust error calculations, with $0.2
  \la \delta L_\text{bol}/L_\text{bol} \la 0.5$.
  Peak luminosities and $^{56}$Ni masses are similarly comparable to previous work.
  SN~2000cb remains an unusual member of this sub-group, owing to the faster rise and flatter plateau than the other supernovae in the sample.
  Initial comparisons with the NLTE atmosphere code \texttt{PHOENIX} show that the direct integration technique re-produces the luminosity of a model supernova spectrum to $\sim$5\% when given synthetic photometry of the spectrum as input.
Our code is publicly   available. The ability to produce bolometric
lightcurves from observed sets of broad-band light curves should be
helpful in the interpretation of other types of supernovae,
particularly those that are not well characterized, such as extremely
luminous supernovae and faint fast objects. 

\end{abstract}

\section{Introduction}

The bolometric luminosity of a supernova is the total radiant luminosity, typically measured in erg s$^{-1}$.
The variation of this luminosity with time after explosion is the lightcurve, and is an important characteristic in the study of transient objects such as supernovae.
Supernovae are, to first order, classified by their spectra, further classification involves the shapes of their lightcurves \citep{filippenko_optical_1997}.
The shapes of lightcurves also reveal important information about a supernova.
Hydrodynamic models of expanding supernova ejecta output bolometric lightcurves which can be compared with those of observed supernovae \citep[for a recent example, see][]{piro_transparent_2014, morozova_light_2015}.
From these comparisons, the model is used to estimate the mass and structure of the progenitor, the total energy of the explosion, and the amount of radioactive $^{56}$Ni synthesized in the supernova.

Determining the properties of a supernova progenitor by matching its observed bolometric lightcurve to one calculated by a hydrodynamic model is predicated on the ability to accurately determine the bolometric lightcurve of a supernova.
For well-observed supernovae with robust distance determinations like SN 1987A, the bolometric luminosity can be determined by integrating the copious photometry directly \citep[see][and references therein]{suntzeff_bolometric_1990}.
For less well-observed supernovae, corrections must be made to the available photometry to account for unobserved radiation from the object.
We seek to analyze these different methods by using a set of supernovae thought to originate from blue supergiant (BSG) progenitors, like the aforementioned SN 1987A.
All of the supernovae have bolometric luminosities published previously in the literature.
We will produce bolometric light curves for each of these supernovae using a variety of techniques from the literature, and characterize any variability that might arise from the use of these different methods. 

\section{Techniques}\label{sec:methods}

The techniques for calculating a bolometric lightcurve from observed photometric magnitudes can be classified into two broad categories: direct integration and bolometric correction.

Direct integration uses only the observed photometry, converting broad-band filter magnitudes to monochromatic fluxes at wavelengths representative of the filters.
These fluxes are then integrated, typically using the trapezoidal method, to generate a value often referred to as the \emph{quasi-bolometric flux}.
This quasi-bolometric flux represents only the observed portion of the total spectral energy distribution of the supernova, and does not include flux which falls blueward or redward of the observable range of the telescope.
The quasi-bolometric flux is typically augmented by UV and IR corrections --- estimates of the missing flux blueward and redward of the observed flux.
Usually, these corrections are made by fitting a blackbody to the observed flux, and integrating that blackbody function from the reddest observed wavelength to infinity, and from the bluest observed wavelength to zero.
Because the SED of supernovae are known to depart significantly from that of a blackbody due to line blanketing in the UV, there are a variety of ways that different groups handle UV corrections, which will be discussed in detail below.

Bolometric correction methods use the bolometric lightcurves of well-observed supernovae (usually calculated using the direct integration technique as mentioned above) to find correlations between an observable quantity such as color and the bolometric correction $BC = m_{bol} - (V - A_V)$.
With this, magnitudes in a filter band can be converted to bolometric magnitudes, and then into bolometric luminosities.
By finding polynomials which describe the relationship between color and bolometric correction, the bolometric luminosity of a less well-observed supernova can be calculated simply my making color observations and a distance estimate.
This assumes, of course, that the same relationship found between the color and bolometric correction of the template supernova exists for the less well-observed supernova.
It is therefore important to use several different well-observed supernovae to establish the polynomials used to transform color into a bolometric correction.

\subsection{Conversion of observed magnitudes to monochromatic fluxes}

The calculation of a bolometric luminosity from observational photometric data begins with the conversion from magnitude to monochromatic flux.
The standard relationship between magnitudes and fluxes for two observed objects is
\begin{equation}
    m_1 - m_2 = -2.5\log_{10}\left(\frac{f_1}{f_2}\right).
    \label{eq:magflux}
\end{equation}
Many photometric systems are characterized by the ``flux at zero magnitude" in each passband --- the monochromatic flux at the effective wavelength of the filter that corresponds to a magnitude of zero.
Values of the flux corresponding to zero magnitude for the UBVRIJHKL Cousins-Glass-Johnson system are reported in \citet{bessell_model_1998} (their Table A2).
In the case that the flux at zero magnitude is available from the literature, the equation to transform from apparent magnitude to flux incident at the top of Earth's atmosphere is straightforward:
\begin{equation}
    f(\lambda_{\text{eff}}) = f_0(\lambda_{\text{eff}}) 10^{-0.4 (m - 0)}
\end{equation}

    Some photometric systems, such as the natural system developed for the Carnegie Supernova Project (CSP) \citep[see][and references within]{stritzinger_carnegie_2011} report zero-points in addition to or instead of $f_0(\lambda_{\text{eff}})$.
The zero-point of filter $X$ is defined in relation to the observed magnitude of a standard star $m_{\text{std}}$ through filter $X$ and the mean flux of the star across the filter bandpass $\bar{f}_X$. The mean flux of the star across the filter bandpass is defined as
\begin{equation}
    \bar{f}_X = \frac{\int_{\lambda_a}^{\lambda_b} f_{\text{std}}(\lambda) S_X(\lambda) d\lambda}{\int_{\lambda_a}^{\lambda_b} S_X(\lambda) d\lambda},
\end{equation}
where $f_{\text{std}}(\lambda)$ is the spectral energy distribution (SED) of the standard star (incident at the top of Earth's atmosphere) in units of erg s$^{-1}$ cm$^{-2}$ \AA$^{-1}$, $S_X(\lambda)$ is the response function of the bandpass $X$, $\lambda_a$ and $\lambda_b$ are the bounds of the filter response function.
The zero point is defined by
\begin{equation}
    m_{\text{std}} = -2.5\log_{10}(\bar{f}_X) + \text{ZP}_X.
\end{equation}

Using this zero point, the equation to transform from apparent magnitude to mean flux incident at the top of Earth's atmosphere is
\begin{equation}
    \bar{f}_X = 10^{-0.4 (m - \text{ZP}_X)}.
    \label{eq:mean_flux_zeropoint}
\end{equation}

Complications to this procedure arise when examining photometry from different photometric systems.
CCD detectors integrate photon counts rather than energy fluxes.
As a result, the zeropoints reported for some photometric systems assume fluxes and bandpasses in photon units, rather than units of energy (for details, see \citet{bessell_model_1998}, Appendix E and \citet{hamuy_distance_2001}, Appendix B).
It should also be noted that the zeropoints of photometric systems such as the CSP natural system are calculated using the integrated flux within the bandpass ($F_X$, in photons cm$^{-2}$ s$^{-1}$) rather than the mean flux over the bandpass ($\bar{f}_X$, in photons cm$^{-2}$ s$^{-1}$ \AA$^{-1}$).
The two are related by
\begin{equation}
    \bar{f}_X = \frac{F_X}{\int_{\lambda_a}^{\lambda_b} S_X(\lambda) d\lambda}.
\end{equation}
For standard photometric systems, transmission functions are available in on-line databases such as the Asiago Database on Photometric Systems (ADPS\footnote{\url{http://ulisse.pd.astro.it/Astro/ADPS/}}) \citep{moro_asiago_2000, fiorucci_asiago_2003}.

\subsection{Direct integration techniques}\label{sec:direct_int}

The fluxes obtained by converting the observed broad-band photometry can be used to directly estimate the bolometric luminosity of a supernova.
This is accomplished by integrating the observed fluxes, and then correcting for un-observed flux that falls outside the wavelength range covered by the filters used in the photometric observations.

\subsubsection{Integration of observed fluxes}\label{subsec:fqbol}

In order to integrate the fluxes over wavelength, each flux must be assigned to a particular wavelength within the bandpass of the filter used to obtain the magnitude measurement.
Many authors use the effective wavelength $\lambda_{\text{eff}}$ of
the filter band as the wavelength of the monochromatic flux
$\lambda_{X}$ for filter $X$ \citep[see, for example,][]{hamuy_sn_1988, stritzinger_optical_2002, pastorello_sn_2005, folatelli_sn_2006, bersten_bolometric_2009}.

With the fluxes assigned to wavelengths, different integration techniques can be used to determine a ``quasi-bolometric" flux $F_{\text{qbol}}$.
Many authors use trapezoidal integration, which integrates the
observed fluxes in each filter band $X$, using linear interpolation
between the observations \citep[see, for example,][]{hamuy_sn_1988, clocchiatti_study_1996, elmhamdi_photometry_2003}

\begin{equation}
    F_{\text{qbol}} = \frac{1}{2} \sum_{X=1}^{N} (\lambda_{X+1} - \lambda_{X})(\bar{f}_{X+1} + \bar{f}_{X}) 
\end{equation}
From this quasi-bolometric flux, the quasi-bolometric luminosity is determined using the usual relation
\begin{equation}
    L_{\text{qbol}} = 4\pi D^2 F_{\text{qbol}}
\label{eq:lqbol_fqbol}
\end{equation}
where $D$ is the distance to the supernova.

\citet{taddia_type_2012} use a different scheme to determine the quasi-bolometric flux. First, a cubic spline curve is fit to the observed epochs, and that curve is then integrated from the shortest to longest value of $\lambda_{X}$.

\subsubsection{Corrections for unobserved flux}

Because the bolometric luminosity of a supernova includes luminosity at wavelengths which fall outside the range of ground-based broad-band photometry, estimates must be made of the missing flux blueward of the shortest wavelength filter and redward of the longest wavelength filter used on a particular observation.

There are different methods used throughout the literature to estimate this missing flux.
\citet{hamuy_sn_1988} estimate the missing flux from SN~1987A in the IR by assuming that the flux from the M band to infinity is equal to the flux observed between the L and M bands, an assumption based on the behavior of a blackbody of temperature \unit[5000]{K}.
Because the observations of SN~1987A extend so far into the IR, the missing flux in the IR using this estimation is below 1\% of the total flux at all observed epochs.

Unlike SN~1987A, most supernovae lack extensive, frequent IR observations over the course of their photometric evolution.
For these supernovae, a different approach must be used.
\citet{patat_metamorphosis_2001} assume a constant IR correction of 35\% in 
their analysis of SN 1998bw, the value being taken from the IR contribution of 
the last available set of JHK observations.
\citet{elmhamdi_photometry_2003} similarly use a constant correction of \unit[0.19]{dex} to scale up the radioactive tail of SN 1999em, provided by integrating the latest observed J, H, and K fluxes.
\citet{schmidt_photometric_1993} used the ratio $L_{\text{VRI}}/L_{\text{bol}}$ of SN~1987A to correct the VRI-derived luminosity of SN~1990E, assuming
\begin{equation}
    L(\text{SN}) = L_{\text{qbol}}(\text{SN}) \times \frac{L_{\text{bol}}(\text{1987A})}{L_{\text{qbol}}(\text{1987A})}.
\label{eq:87A_scaling}
\end{equation}
This approximate bolometric correction factor is valid only if both objects undergo identical evolution in color \citep{pastorello_sn_2012}.
This assumption is reasonable enough to provide an estimate of the bolometric luminosity, and was also used by \citet{clocchiatti_study_1996} in the study of SN~1992H and more recently by \citet{pastorello_sn_2012} in the study of SN~2009E.

Most recent papers correct for missing flux in the IR by fitting a blackbody SED to the reddest observed fluxes (typically near-IR filters,) and integrating the blackbody flux redward of the longest effective wavelength to $\lambda = \infty$ \citep{folatelli_sn_2006, bersten_bolometric_2009, lyman_bolometric_2014}.

The techniques used to correct for missing flux in the UV depend on the age of the supernova and the details of the SED.
The primary cause of this diversity stems from the fact that in the UV, supernovae depart rapidly from the idealized blackbody SED as they expand and cool.
The departure is most pronounced during the plateau and radioactive
decline phases of a Type II supernova lightcurve, and is due to the
formation of a dense forest of absorption lines in the UV from
iron-peak elements in the ejecta \citep[see, for example,][]{baron_non-local_1996}

Early in the visible evolution of a supernova, the effective temperature is on the order of $T_\text{eff}\sim\unit[1\times10^4]{K}$ and lacks strong features, closely approximating the spectrum of a high-temperature blackbody \citep{filippenko_optical_1997}.
The peak of the emission is then $\sim \unit[2900]{\AA}$, inaccessible
to ground-based telescopes as the theoretical UV cutoff for most
observatories lies at $\lambda = \unit[3000]{\AA}$ \citep[see, for example,][]{nitschelm_theoretical_1988}.
At early times, therefore, the UV correction factor will be significant.

\citet{bersten_bolometric_2009} found that for a sample of three well-observed supernovae (which included SN 1987A) the missing flux in the UV at early times was as high as 50\% of the total observed flux.
Using supernova atmosphere models, they found even higher UV corrections of up to 80\% when $B-V \lesssim 0.2$.

Space-based UV photometry of supernovae is available from orbital instruments such as the \textit{Swift} UV-Optical Telescope (UVOT) \citep[see][]{roming_swift_2005}.
Because \textit{Swift} UVOT observations reach farther into the UV than ground-based observations, the need to correct for unobserved UV light will decrease.
In a study of \textit{Swift} photometry of 50 core-collapse supernovae, \citet{pritchard_bolometric_2014} find that UV corrections drop to $\sim 10\% - 30\%$ at early times.

The methods used to calculate the UV correction fall largely into two categories: black-body extrapolation and linear extrapolation.
As mentioned above, at early times the spectrum of a supernova is largely similar to that of a high-temperature blackbody.
To correct for the large amount of flux emitted in the UV at these times, many past studies have fit a blackbody curve to the monochromatic fluxes derived from broad-band photometry, and integrated under the blackbody curve from $\lambda = 0$ to a wavelength near the shortest observed filter.
\citet{bersten_bolometric_2009} use the effective wavelength of the U filter as the end-point of the integration, while \citet{lyman_bolometric_2014} use the blue edge of the of the U band.

As the supernova cools and expands, atmosphere models diverge from a simple Planck function in the UV.
Dense forests of absorption lines from Fe-group elements cause short-wavelength U and B filter observations to drop below the magnitudes expected from a blackbody curve of the same effective temperature.
To correct for unobserved UV flux in these epochs, it is common to augment the observed fluxes with a linear extrapolation from a characteristic wavelength where the flux is assumed to drop to zero ($\lambda = \unit[2000]{\AA}$ in \citet{bersten_bolometric_2009} and \citet{lyman_bolometric_2014}, and $\lambda = \unit[3000]{\AA}$ in \citet{folatelli_sn_2006}) to the effective wavelength of the shortest observed filter.

\subsection{Bolometric correction techniques}\label{sec:BC}

The techniques used to calculate the bolometric luminosity in \autoref{sec:direct_int} work best for supernovae observed over their entire evolution with frequent observations in multiple bandpasses covering the widest possible wavelength range.
Due to the realities of observing transient objects, there are very few supernovae which match those criteria.
We therefore expect that supernovae with less frequent observations using limited filtersets will have larger uncertainties in the bolometric luminosities derived using direct integration techniques.

However, it is possible to leverage the plentiful observations of supernovae like SN~1999em and SN~1987A to help determine the bolometric luminosity of a less well-observed supernova.
This can be accomplished by first determining the bolometric luminosities of the well-observed ``template'' supernovae and then computing bolometric correction factors, typically the difference between the apparent bolometric magnitude and the apparent V-band magnitude

\begin{equation}
  \label{eq:bc}
  BC = m_\text{bol} - (V - A_\text{V, TOT}).
\end{equation}

As shown by \citet{hamuy_type_2001}, the values of the bolometric correction for template supernovae SN~1999em and SN~2000cb correlate strongly with the intrinsic $BVI$ colors of the supernovae \citep[Figure 5.3 in][]{hamuy_type_2001}.
The correlation can be quantified by a polynomial fit of the form

\begin{equation}
  \label{eq:bc_color}
  BC(\text{color}) = \sum_{i = 0}^{n} c_i (\text{color})^i
\end{equation}

where $n$, the order of the fit, varies with the chosen color.
Later work by \citet{bersten_bolometric_2009} refined the polynomial fits by adding data from the well-observed template supernova SN 2003hn (coefficients $c_i$ given in their Table 1.)
Additional work by \citet{lyman_bolometric_2014} using a sample of 21 well-observed template supernovae from the literature produced a third set of second-order polynomial coefficients and expanded the range of valid color combinations to include those from the set $BgVriI$ (their Tables 1, 2, and 3.)

\citet{pejcha_global_2015} developed a theoretical model of supernova lightcurve evolution and performed a large-scale least-squares fit using data from 26 supernovae.
With their results, they calculate bolometric corrections using a broader range of filters than previous studies, and produce a fourth set of fifth-order polynomial coefficients (their Table 8.)

With the polynomial coefficients $c_i$ from one of the above datasets, it is possible to calculate the bolometric correction to convert an observed magnitude to a bolometric magnitude using only two-filter colors of the observed supernova.
The resulting bolometric magnitudes can then be converted into bolometric luminosities after choosing an appropriate bolometric magnitude zeropoint ($\text{ZP} = -11.64$ in \citet{bersten_bolometric_2009}, but appears with the wrong sign in their Equation 4.)

\begin{equation}
    \label{eq:lbol_bc}
    \log_{10}(L_\text{bol}) = -0.4[BC(\text{color}) + V - A_\text{V} - \text{ZP}] + \log_{10}(4\pi D^2)
\end{equation}

With the ability to calculate a quasi-bolometric luminosity $L_\text{qbol}$, a multitude of techniques for correcting that luminosity for unobserved UV and IR light to calculate the direct integration luminosity $L_\text{D}$ and multiple sets of coefficients for determining the luminosity from bolometric corrections $L_\text{BC}$, we now turn our attention to the sample of SN~1987A-like supernovae sometimes referred to as ``peculiar'' Type~II-P.

\section{Supernovae}\label{sec:sne}

The supernovae used in this sample were selected by their common origin with SN 1987A --- all are thought to originate from compact blue supergiant progenitors.
Our sample includes SN 1998A, SN 2000cb, SN 2006V, SN 2006au, and SN 2009E. These are the best observed of this class of supernovae (except, of course, for SN 1987A.)
In the sections to follow, we will use the techniques discussed above to determine the bolometric lightcurves of these supernovae, and compare them with those previously published in the literature.

\subsection{SN~1998A}\label{sec:98a_obs}

SN~1998A was discovered on 1998 January 6.77 UT by \citet{williams_supernova_1998} as part of the automated supernova search by the Perth Astronomy Research group.
Follow-up observations determined the location to be $\alpha = 11^\text{h}09^\text{m}50^\text{s}.33$, $\delta = -23^{\circ}43'43''.1$ (J2000).
Preliminary photometric data from the Perth Astronomy Research Group was published by \citet{woodings_light_1998}.
A more detailed analysis of the object was published by \citet{pastorello_sn_2005}, including spectroscopy, atmospheric models, and a pseudo-bolometric light curve.

The distance, extinction, and explosion date for SN~1998A are taken
from \citet{pastorello_sn_2005} and are compiled in \autoref{tab:sn_parameters}.
The distance modulus of $\mu = 32.41$ to the supernova was taken from the recession velocity of the host galaxy IC 2627 found in LEDA\footnote{\url{http://leda.univ-lyon1.fr}} \citep{makarov_hyperleda._2014}, assuming $H_0 = \unit[65]{km\; s^{-1}\;Mpc^{-1}}$.
This corresponds to a distance of $D_{\text{98A}} = \unit[30 \pm 7]{Mpc}$.
Foreground extinction of $A_{\text{B, TOT}} = 0.52$ from \citet{schlegel_maps_1998} is the only significant source of extinction along the line-of-sight to the supernova.
No host galaxy extinction is assumed, evidenced by the lack of narrow Na~I D absorption in the spectrum of the supernova \cite{leonard_spectropolarimetry_2001}.
The explosion date is taken to be JD $245080\pm4$ from \citet{woodings_light_1998} from shifting the shape of the light curve to match that of SN 1987A.

The bolometric lightcurve presented in \citet{pastorello_sn_2005} is a pseudo-bolometric lightcurve, meaning that no corrections were made for unobserved flux in the UV or IR.

\subsection{SN~2000cb}\label{sec:00cb_obs}

SN~2000cb was discovered on 2000 April 27.4 UT by \citet{papenkova_supernova_2000} as part of the Lick Observatory Supernova Search.
The supernova was located at $\alpha = 16^\text{h}01^\text{m}32^\text{s}.15$, $\delta = +1^{\circ}42'23''.0$ (J2000) in the spiral galaxy IC 1158.
Photometry of the object has been published by \citet{hamuy_type_2001} and \citet{kleiser_peculiar_2011}.

The supernova parameters given by \citet{kleiser_peculiar_2011}, are shown in \autoref{tab:sn_parameters}.
The distance $D_{00cb} = \unit[30\pm7]{Mpc}$ is taken from the SFI++ dataset of Tully-Fisher distances published by \citet{springob_erratum:_2009}.
This distance is slightly lower than the range of $\unit[\sim31-37]{Mpc}$ given by the various expanding photosphere method analyses in \citet{hamuy_type_2001}.
The host galaxy reddening of the supernova is minimal --- as in the case of SN~1998A the supernova spectrum shows no evidence of narrow Na~I D absorption.
We adopt the Galactic extinction used in \citet{kleiser_peculiar_2011} of $E(B-V)_{\text{Gal}} = \unit[0.114]{mag}$ from \citet{schlegel_maps_1998} with the \citet{cardelli_relationship_1989} slope of $R_V = 3.1$ to give $A_{\text{V, TOT}} = 0.373$.
The explosion date of JD $2451656 \pm 4$ is calculated by \citet{kleiser_peculiar_2011} using a cubic spline extrapolation of their first five unfiltered observations of SN~2000cb.
\citet{hamuy_type_2001} arrives at a similar value of $t_0 = 2451653.8$ using an average of six expanding photosphere method solutions.

Published bolometric lightcurves of SN~2000cb appear in both \citet{hamuy_type_2001} and \citet{kleiser_peculiar_2011}.
The lightcurve published by \citet{hamuy_type_2001} is found using a bolometric correction technique very similar to the more recent one published by \citet{bersten_bolometric_2009}.
The lightcurve published by \citet{kleiser_peculiar_2011} is calculated by fitting a blackbody spectrum to the observed optical photometry of SN~2000cb, leaving the luminosity, radius, and temperature as free parameters.

\subsection{SN~2006V}\label{sec:06v_obs}

SN~2006V was discovered by \citet{chen_supernova_2006} on 2006 February 4.67 UT as part of the Taiwan Supernova Survey.
The supernova was located at $\alpha = 11^\text{h}31^\text{m}30^\text{s}.01$, $\delta = -2^\circ17'52''.2$ (J2000) in the spiral galaxy UGC~6510.
Photometry of the object was published by \citet{taddia_type_2012} as part of the Carnegie Supernova Project.

We adopt basic supernova parameters first published in the analysis of SN~2006V undertaken by \citet{taddia_type_2012}:
Using the Hubble constant of $H_0 = \unit[73.8\pm2.4]{km \; s^{-1} \; Mpc^{-1}}$ from \citet{riess_3_2011} and the measured redshift of the supernova of $z = 0.0157 \pm 0.0013$, the distance is calculated to be $\unit[72.7 \pm 5]{Mpc}$.
The lack of Na~I D absorption in the spectra of SN~2006V lead \citet{taddia_type_2012} to rule out significant host galaxy extinction, and use the NED\footnote{\url{https://ned.ipac.caltech.edu/}} value of $E(B-V)_{\text{Gal}} = 0.029$ from \citet{schlegel_maps_1998} with the \citet{cardelli_relationship_1989} slope of $R_v = 3.1$ to give $A_{\text{V, TOT}} = 0.09$.
\citet{taddia_type_2012} use EPM analysis (outlined in their section 5.1) to constrain the explosion epoch of SN~2006V to JD $2453748 \pm 4$.

The bolometric lightcurve of SN~2006V has been published previously in \citet{taddia_type_2012} (their Figure 14). As mentioned in \autoref{subsec:fqbol}, this bolometric luminosity was calculated using a modified direct integration technique, where observed magnitudes were first converted to fluxes, and then a cubic spline function was fit to the flux points.
This cubic polynomial was then integrated over wavelength, with corrections made for IR flux by integrating a Rayleigh-Jeans tail redward of the $H$ band and Wien tail blueward of the $u$ band.

\subsection{SN~2006au}\label{sec:06au_obs}

SN~2006au was discovered on 2006 March 7.2 by \citet{trondal_supernova_2006} as part of the Tenagra Observatory Supernova Search.
The supernova was located at $\alpha = 17^\text{h}57^\text{m}13^\text{s}.56$, $\delta = +12^\circ11'03''.2$ (J2000) in the spiral galaxy UGC 11057.
Photometry of the supernova was first published in \citet{taddia_type_2012}, along with the data from SN~2006V.

The distance, reddening, and explosion date are taken from \citet{taddia_type_2012}.
Using the same techniques as described in \autoref{sec:06v_obs} the distance to SN~2006au was found to be $\unit[46.2 \pm 3.2]{Mpc}$.
\citet{taddia_type_2012} find clear Na~I D absorption lines in the spectrum of SN~2006au with an equivalent width of $\unit[0.88 \pm 0.11]{\AA}$, and a corresponding host galaxy color excess of $E(B-V)_{\text{host}} = 0.141$ using the correlation between Na~I D equivalent width and host galaxy reddening published in \citet{turatto__2003}.
Combining this value with the \citet{schlegel_maps_1998} value of galactic color excess $E(B-V)_\text{Gal} = 0.172$ gives $A_\text{V, TOT} = 0.97$.
Using EPM estimates, \citet{taddia_type_2012} constrain the explosion date to JD $2453794 \pm 9$.

The bolometric lightcurve of SN~2006au was calculated in the same way as that of SN~2006V.

\subsection{SN~2009E}\label{sec:09e_obs}

SN~2009E was discovered on 2009 January 3.06 UT by \cite{boles_supernova_2009} in the spiral galaxy NGC 4141, at location $\alpha = 03^\text{h}54^\text{m}22^\text{s}.83$ and $\delta = -19^\circ10'54''.2$.
\cite{prosperi_supernova_2009} noted that observations of the supernova on 2009 March 8.05 UT show that it has brightened by a full magnitude over the course of a month.
Follow-up spectroscopy on 2009 March 24.88 UT by \citet{navasardyan_supernova_2009} revealed SN~2009E to be a type II supernova with strong barium features analogous to SN~1987A.

Photometry of the object was first published by \citet{pastorello_sn_2012}.
For the first three months of observation, the object was monitored largely by amateur astronomers \citep[for details on the reduction and calibration of unfiltered amateur images, see][]{pastorello_sn_2012}.
The observational parameters of the supernova are given in \autoref{tab:sn_parameters}.
The explosion date of SN~2012E was found to be JD $2454832.5_{-5}^{+2}$ by comparing the early photometric evolution to that of SN~1987A.
The distance to the host galaxy was calculated to be $\unit[29.97 \pm 2.10]{Mpc}$ using the redshift and a Hubble constant of $H_0 = 72 \pm 5$ km s$^{-1}$ Mpc$^{-1}$.
The galactic extinction from \citet{schlegel_maps_1998} in the direction of SN~2009E is $E(B-V)_\text{Gal} = 0.02$.
Evidence of faint Na~I D absorption in the spectrum suggests a host galaxy extinction of $E(B-V)_\text{Host} = 0.02$ using the relation of \citet{turatto__2003}, for a total extinction of $E(B-V)_\text{tot} = 0.04$. With the \citet{cardelli_relationship_1989} slope of $R_V = 3.1$, the value of $A_\text{V, TOT} = 0.124$.

The bolometric lightcurve of SN~2009E was first reported in \citet{pastorello_sn_2012}.
The quasi-bolometric lightcurve was found by converting the observed photometry to monochromatic flux, then integrating over wavelength.
Corrections for unobserved flux were made by assuming the color evolution was identical to SN~1987A, and scaling the quasi-bolometric flux of SN~2009E by the ratio of the bolometric to quasi-bolometric flux of SN~1987A (see \autoref{eq:87A_scaling})

\section{Implementation}

As can be seen in \autoref{sec:sne}, a wide variety of techniques have been used in the literature to determine the bolometric luminosities of supernovae with BSG progenitors.
This itself is a subset of the techniques which have been used to calculate the bolometric luminosities of supernovae, in general --- some of which were discussed in \autoref{sec:methods}.

In order to understand the results obtained by each of these techniques, we have written the \underline{Super}nova \underline{Bo}lometric \underline{L}ightcurves code (\texttt{SuperBoL}).
The goal of this code is to provide the community with a set of standardized tools to calculate $L_\text{bol}$, and to provide best estimates of the uncertainties in those calculations. 
The python code is open-source and available on GitHub\footnote{\url{https://github.com/JALusk/SuperBoL}}.
This repository includes the photometry of the supernovae studied in this paper in an HDF5 file, making it possible to re-produce the results detailed in \autoref{sec:results}.
We welcome contributions, corrections, and feature requests from the community.

\subsection{Direct integration in SuperBoL}\label{sec:direct}
\texttt{SuperBoL} implements a direct integration technique based on the one published in \cite{bersten_bolometric_2009}.
The observed magnitudes for a supernova are taken from the literature sources cited in \autoref{sec:sne} and converted to fluxes at the effective wavelengths of their filters.
Those fluxes are then dereddened using the reddening law of \cite{cardelli_relationship_1989} with values of $A_\text{V, TOT}$ from \autoref{tab:sn_parameters} using the \texttt{extinction.reddening} function from the Astropy-affiliated package \texttt{specutils}\footnote{\url{https://github.com/astropy/specutils}}.

The bolometric luminosity is then calculated for JD on which at least 4 bandpasses were used in observations of the supernova.
No attempts are made to interpolate the magnitudes of missing bandpasses.

The quasi-bolometric flux $F_\text{qbol}$ is calculated by the trapezoidal integration technique detailed in \autoref{subsec:fqbol} using the \texttt{trapz} routine in the \texttt{numpy} package\footnote{\url{http://www.numpy.org/}}.
In order to estimate missing flux in the IR ($F_\text{IR}$), a blackbody spectrum is fit to the observed fluxes using the \texttt{curve\_fit} function from the python package \texttt{scipy}\footnote{\url{http://www.scipy.org}} \citep{jones_scipy:_2001}.
The best-fit blackbody curve is then integrated from the longest observed effective wavelength to $\lambda = \infty$.
Missing flux in the UV ($F_\text{UV}$) is handled differently depending on the quality of the blackbody fit.
To best re-create the method described in \citet{bersten_bolometric_2009}, the blackbody fit is again integrated, this time from the shortest observed wavelength to $\lambda = 0$, unless the $U$-filter flux falls below that blackbody fit, in which case a linear function linking the shortest observed flux and $f_\lambda = 0$ at $\lambda = \unit[2000]{\AA}$ is integrated instead.
If no $U$-filter observations are available, the integrated blackbody flux is used to calculate $F_{\text{UV}}$

The integration of the Planck function proved unreliable with standard integration packages such as \texttt{integrate.quad} from \texttt{scipy}.
To overcome this difficulty, we express the discrete integral of the Planck function as an infinite series as shown in \autoref{eq:planck_integral} \citep{michels_planck_1968,widger_integration_1976}:

\begin{equation}
  \int_{0}^{\lambda_1} B_{\lambda}(\lambda, T) \; d\lambda = \frac{C_1 T^4}{C_2^4} \sum_{n = 1}^{\infty}\left(\frac{x_1^3}{n} +
    \frac{3x_1^2}{n^2} + \right. \left. \frac{6x_1}{n^3} + \frac{6}{n^4}\right) e^{-nx_1}
  \label{eq:planck_integral}
\end{equation}
where $C_1 = 2hc^2$ and $C_2 = hc/k_B$ are the first and second radiation constants, and $x_1 = C_2/\lambda_1 T$.
This infinite series is truncated in \texttt{SuperBoL} to produce
accuracies to ten digits.
The rate of convergence in the series depends upon the value of $x_1$, and in order to achieve this accuracy we found
we had to include $n = \text{min}(2 + 20/x_1, 512)$ terms in the series\footnote{\url{http://www.spectralcalc.com/blackbody/inband\_radiance.html}}.
Calculating the IR correction is accomplished in the following manner:
\begin{equation}
  \int_{\lambda_2}^{\infty} B_{\lambda}(\lambda, T) \; d\lambda = \int_{0}^{\infty}B_{\lambda}(\lambda, T) \; d\lambda - \int_{0}^{\lambda_2} B_{\lambda}(\lambda, T) \; d\lambda
\end{equation}
where the first term is given by the Stefan-Boltzmann law as $\sigma T^4$, and the second term is calculated using the series solution above.

The three fluxes, $F_\text{qbol}$, $F_\text{IR}$ and $F_\text{UV}$ are added together to form the bolometric flux $F_\text{bol}$.
This flux is then converted to a luminosity using the standard relation
\begin{equation}
L_\text{bol} = 4\pi D^2 F_\text{bol},
\end{equation}
with $D$ taken from \autoref{tab:sn_parameters}.

In the figures to follow, the quasi-bolometric luminosity $L_\text{qbol}$ is calculated simply by excluding any corrections for missing flux in the UV and IR and using $F_\text{qbol}$ to calculate a luminosity using \autoref{eq:lqbol_fqbol}.

\subsection{Bolometric correction in SuperBoL}

We chose to use the set of polynomial coefficients $c_i$ in \autoref{eq:bc_color} from \citet{bersten_bolometric_2009} for this paper, although in the future \texttt{SuperBoL} will be modified to calculate the bolometric luminosity using any set of coefficients and two-filter colors.
\citet{pejcha_global_2015} have shown (their Figure 16) that the results obtained when using the different sets of coefficients outlined in \autoref{sec:BC} are broadly similar, especially over the range of colors typical of our sample of peculiar Type II-P supernovae ($0.5 \lesssim B-V \lesssim 1.5$, $0.5 \lesssim V-I \lesssim 1.0$.)

We have chosen to average together the results from calculating the bolometric luminosity based on the bolometric correction obtained from $B-V$, $V-I$, and $B-I$ colors.
\texttt{SuperBoL} calculates luminosities based on as many of the color combinations as possible on a given JD before averaging.

\subsection{Propagation of uncertainties in SuperBoL}

One of the goals of producing lightcurves with \texttt{SuperBoL} is to estimate the uncertainty in our final bolometric luminosities.
This can then be used to constrain the uncertainties in progenitor properties found by matching theoretical lightcurves to observed ones.

The basic input data to \texttt{SuperBoL} includes the observed photometry $m_X$, the explosion date $t_0$, the total visual extinction $A_\text{V, TOT}$, and the distance $D$.
Uncertainties in the distance and explosion date are shown in \autoref{tab:sn_parameters}, while uncertainties in the observed photometry for each SN are provided in the sources referenced in \autoref{sec:sne}.
All of these measurements and uncertainties are included in the HDF5 file used as a database by \texttt{SuperBoL}.

We have used the general formula for error propagation \citep[see, for
example,][Eq. 3.47]{taylor_introduction_1997} in our code to ensure that uncertainties in the basic input data are reflected in the final luminosity.
For certain numerical operations, this is not so straightforward.
As an example, we must fit a blackbody curve to the observed fluxes.
We compute one standard deviation uncertainties on the fitted parameters by calculating the square root of the diagonal terms in the covariance matrix output by \texttt{curve\_fit}.
Those standard deviations are then used as the uncertainties in those parameters for the purposes of propagation.
We also assume that the RMS dispersions listed in Table 1 of \citet{bersten_bolometric_2009} represent the standard deviation $\sigma_{BC}$ of the bolometric correction $BC(\text{color})$ in our uncertainty propagation.
Uncertainties in the UV and IR correction due to the temperature can be calculated by taking derivatives of the series solution given in \autoref{eq:planck_integral} with respect to the temperature.
These are included in our uncertainty calculations.

\subsection{Testing and Validation of SuperBoL}

As part of the testing and validation of our code, we have written unit tests using the python \texttt{unittest} package to ensure the base functions of the code produce expected results.

This does not ensure, however, that the overall results of the code are correct -- merely that the individual functions produce expected results given certain inputs.
In order to test that the code functions correctly, we need to compare the luminosities produced by \texttt{SuperBoL} with known good luminosities of astrophysical origin.
We do this in two ways --- first with synthetic photometry from blackbody models, and then with synthetic photometry from the pure helium white dwarf (WD) atmosphere models of \citet{holberg_calibration_2006, kowalski_found:_2006, tremblay_improved_2011} and \citet{bergeron_comprehensive_2011}, available online\footnote{\url{http://www.astro.umontreal.ca/~bergeron/CoolingModels}}. The grid of blackbody models was generated to have the same temperatures and luminosities as the white dwarf models, and synthetic photometry was performed to determine absolute magnitudes in the $UBVRIJHK$ and $ugriz$ bands.

\begin{figure}
\centering
  \includegraphics[scale=0.6]{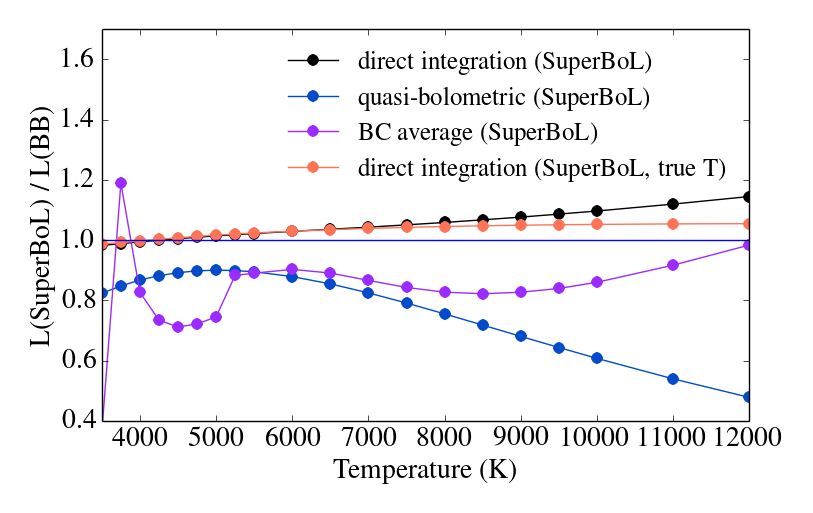}
  \label{fig:bb}
  \caption{Ratio of \texttt{SuperBoL}-derived luminosity to blackbody model luminosity over a temperature range typical in the observed evolution of Type II-P supernovae.}
\end{figure}

In \autoref{fig:bb}, we compare the \texttt{SuperBoL}-derived luminosities with the luminosities of the blackbody models.
Here, the importance of making the UV and IR corrections correctly can be seen.
The quasi-bolometric luminosity underestimates the true luminosity of the blackbody by 10-20\% over the range of temperatures typical of the plateau phase of a Type II-P supernova.
The disagreement increases as the temperature increases, due to an increasing fraction of the total flux shifting blueward of the bluest filter.

The direct integration scheme in \texttt{SuperBoL} performs much better, recovering the blackbody luminosity from synthetic photometry to within 3\% over the range of \unit[3500]{K} - \unit[6000]{K}.
Because the nature of the underlying spectrum is known, the UV correction was made using the blackbody fit integrated to $\lambda = 0$ rather than the linear interpolation detailed in \autoref{sec:direct}.

\begin{figure}
\centering
  \includegraphics[scale=0.6]{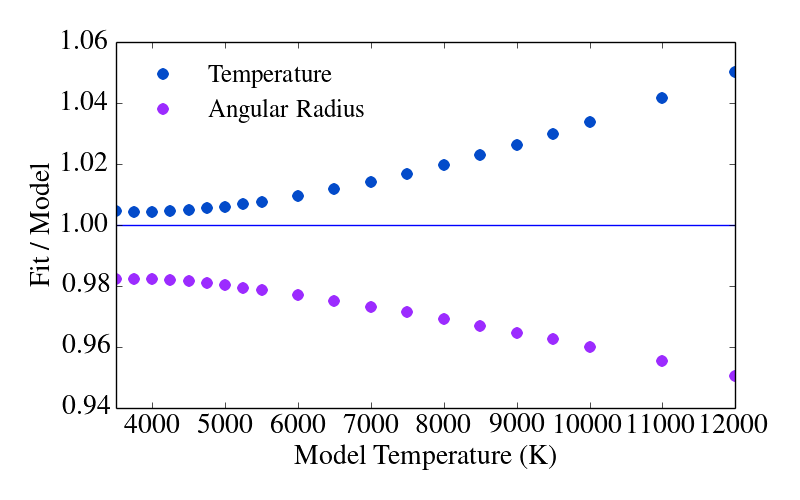}
  \label{fig:temp_theta}
  \caption{Ratio of \texttt{SuperBoL}-derived temperatures and angular radii to those of the blackbody models underlying the synthetic photometry used to validate the code.}
\end{figure}

The gap between the true luminosity and the direct integration luminosity from \texttt{SuperBoL} widens as the temperature increases, partially due to mismatches between the true temperature of the blackbody model underlying the synthetic photometry and the temperature recovered from fitting a blackbody to the fluxes derived from those magnitudes.
The mismatch is shown in \autoref{fig:temp_theta}, which reveals that using \texttt{curve\_fit} to determine the temperatures and angular radii slightly overestimates the temperatures and underestimates the angular radii of the blackbody models from which the synthetic photometry was taken.
To show the extent to which these mismatches affect the luminosities, we re-computed the UV and IR corrections using the true temperatures and angular radii from our blackbody models rather than the ones found by fitting the monochromatic fluxes.
As shown in \autoref{fig:bb}, This brings the luminosities derived from the high-temperature blackbody photometry to within 6\% of the true luminosities.

\begin{figure}
\centering
  \includegraphics[scale=0.6]{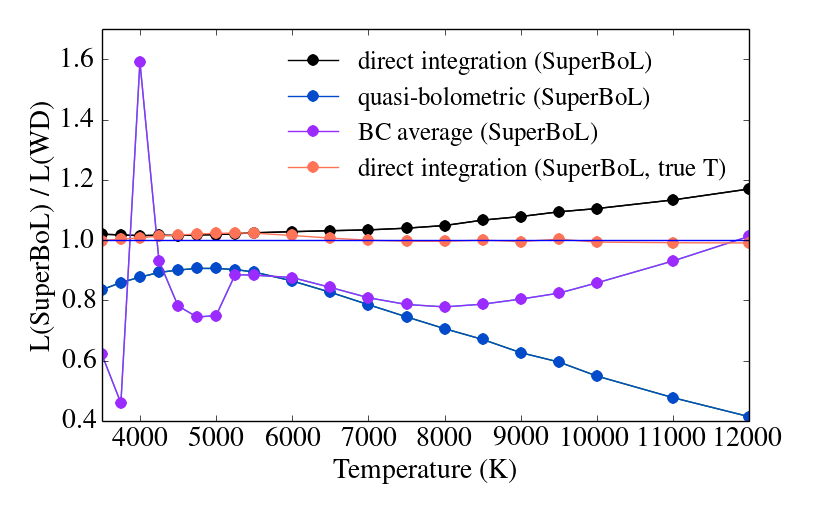}
  \label{fig:wd}
  \caption{Ratio of \texttt{SuperBoL}-derived luminosity to white dwarf model luminosity over a temperature range typical in the observed evolution of Type II-P supernovae.}
\end{figure}

A comparison of the \texttt{SuperBoL}-derived luminosities with the luminosities calculated from the bolometric magnitudes given in the WD models is shown in \autoref{fig:wd}.
The same methods were used in both cases, and the UV correction was again made using the full UV tail of the blackbody function rather than the linear interpolation.
The WD model results closely resemble those of the blackbody models, and luminosities are recovered to within $\sim$3\% over the range typical of the plateau-phase of Type II-P supernovae.
Using the true temperatures, \texttt{SuperBoL} recovers the luminosity to within $\sim$2\% over the temperature range \unit[3500]{K} --- \unit[6000]{K}.

\section{Results}\label{sec:results}

\subsection{Supernova Parameters}\label{sec:sn_parameters}
In order to compare our bolometric lightcurves with those published elsewhere in the literature, we have adopted the same explosion dates, extinction estimates, and distances used in previous studies of each supernova.
These parameters are listed in \autoref{tab:sn_parameters}, along with references to the literature.
Due to this choice, the bolometric lightcurves of our objects are not calculated using a consistent distance scale.
Inter-comparisons between objects must be made with this in mind, since some of the distances from the literature are estimated from redshifts using values of $H_0$ which vary from $H_0 = 73.8 \pm 2.4$ km s$^{-1}$ Mpc$^{-1}$ in the cases of SN~2006V and SN~2006au (\autoref{sec:06v_obs} and \autoref{sec:06au_obs}) to $H_0 = 65$ km s$^{-1}$ Mpc$^{-1}$ in the case of SN 1998A (\autoref{sec:98a_obs}).

\subsection{Lightcurves}\label{sec:lightcurves}

The bolometric lightcurves produced by \texttt{SuperBoL} reveal the large variation in luminosities which results from the different techniques used in the literature.
Unsurprisingly, the quasi-bolometric technique results in systematically lower luminosities than any of the other methods.
The quasi-bolometric technique is also dependent upon the number of filters used in the calculation, since no corrections are made for the unobserved flux which falls outside the wavelength range of the filters used.
The results of the direct integration and bolometric correction methods agree in most cases with previously published lightcurves in the literature, with the direct integration technique producing systematically higher luminosities in the cases of SN~1998A, SN~2006V, and SN~2006au.
Our results also show the high degree of uncertainty in luminosity calculations, typically in the range of $\pm$15\% for supernovae with low uncertainties in their distances to $\pm$50\% for supernovae with significant distance uncertainties.

\begin{figure}
\centering
    \includegraphics[scale=0.6]{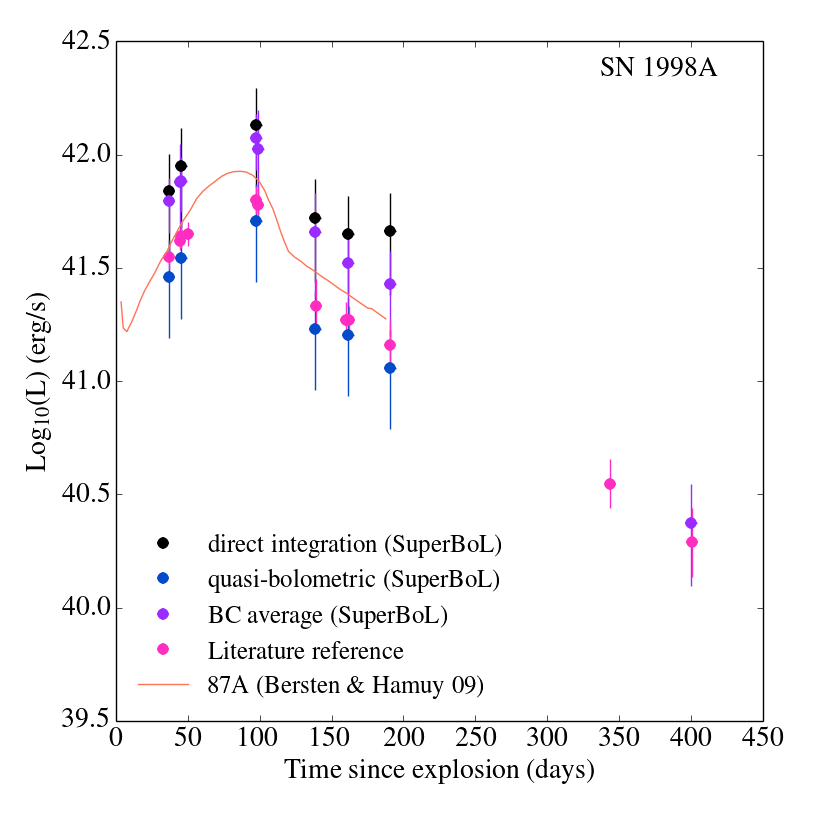}
    \label{fig:98a_lc}
        \caption{Bolometric lightcurves of SN 1998A. The previously published lightcurve is that of \cite{pastorello_sn_2005}, described in \autoref{sec:98a_obs}}
\end{figure}

\begin{figure}
\centering
    \includegraphics[scale=0.6]{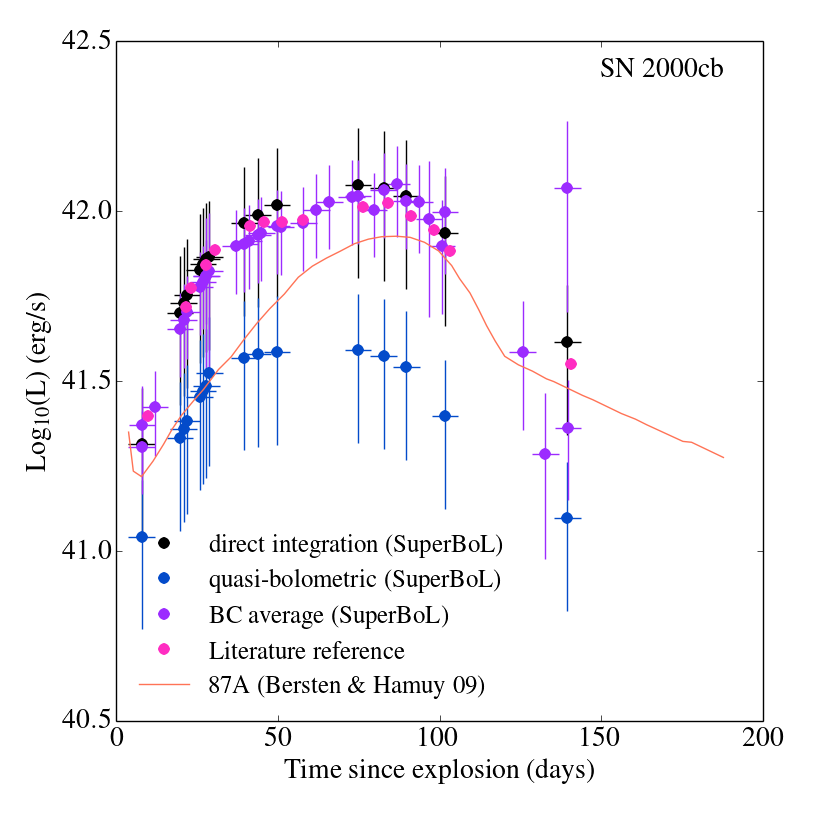}
    \label{fig:00cb_lc}
        \caption{Bolometric lightcurves of SN 2000cb. The previously published lightcurve is that of \cite{hamuy_type_2001}, described in \autoref{sec:00cb_obs}}
\end{figure}

\begin{figure}
\centering
    \includegraphics[scale=0.6]{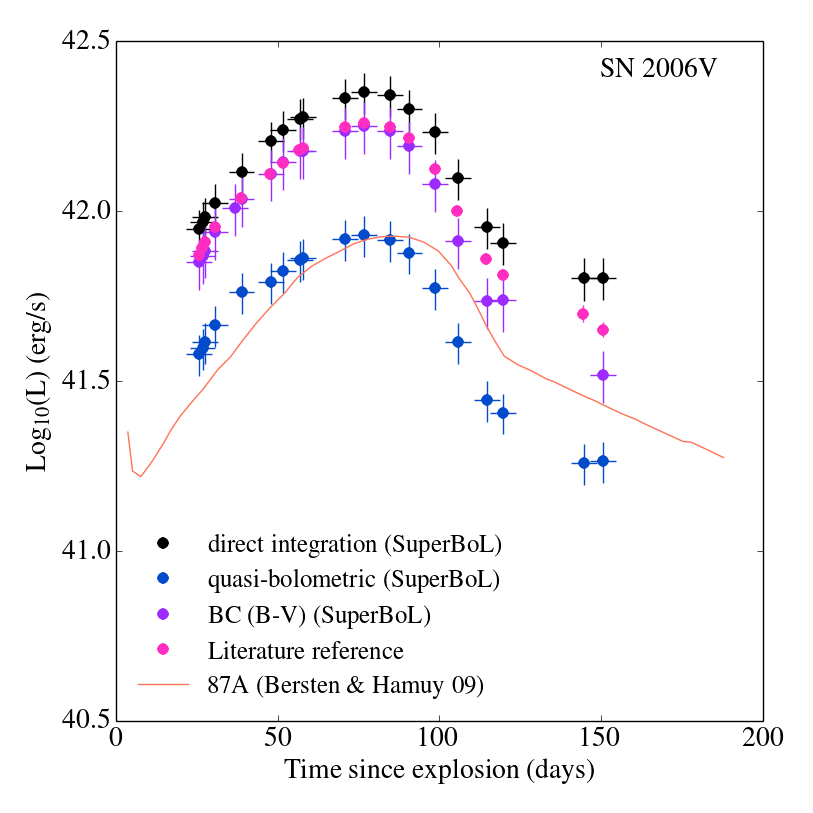}
    \label{fig:06v_lc}
        \caption{Bolometric lightcurves of SN 2006V. The previously published lightcurve is that of \cite{taddia_type_2012}, described in \autoref{sec:06v_obs}}
\end{figure}

\begin{figure}
\centering
    \includegraphics[scale=0.6]{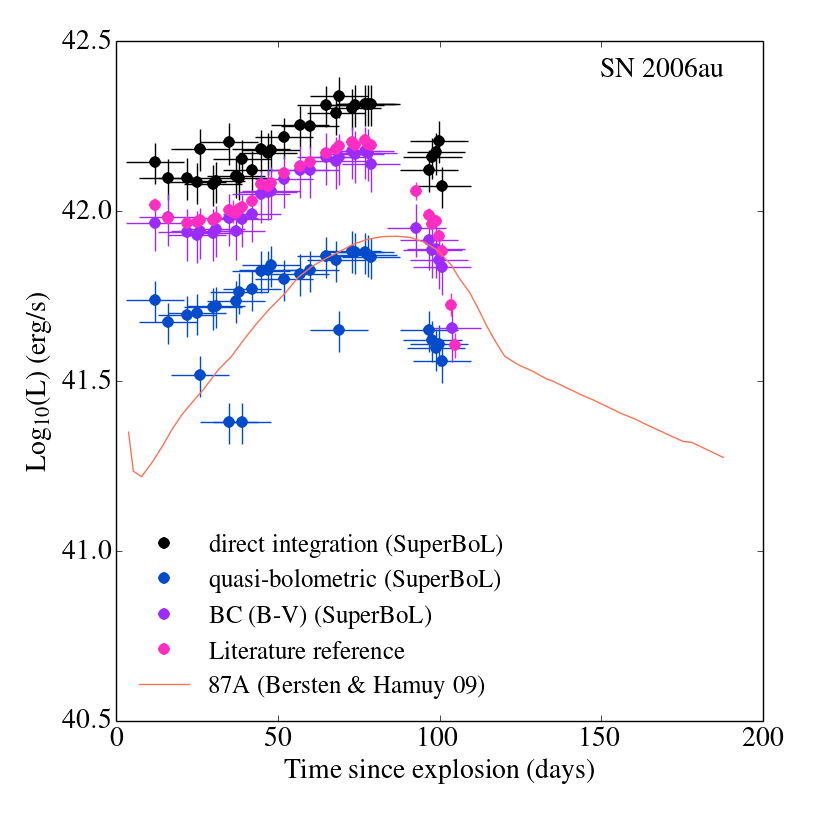}
    \label{fig:06au_lc}
        \caption{Bolometric lightcurves of SN 2006au. The previously published lightcurve is that of \cite{taddia_type_2012}, described in \autoref{sec:06au_obs}}
\end{figure}

\begin{figure}
\centering
    \includegraphics[scale=0.6]{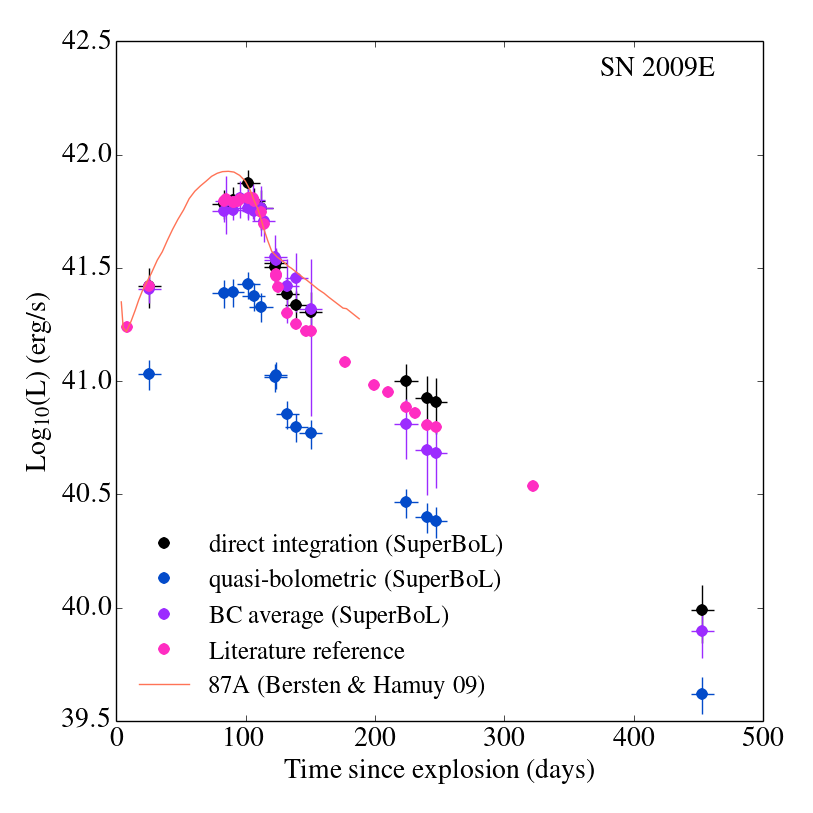}
    \label{fig:09e_lc}
        \caption{Bolometric lightcurves of SN 2009E. The previously published lightcurve is that of \cite{pastorello_sn_2012}, described in \autoref{sec:09e_obs}}
\end{figure}

The lightcurves shown in Figures~\ref{fig:98a_lc} -- \ref{fig:09e_lc} include the absolute errors calculated by \texttt{SuperBoL}, with an upper error of $\log(L + \delta L)$ and a lower error of $\log(L - \delta L)$.
This results in asymmetric errorbars on the plot, since we have adopted the common practice of plotting the logarithm of the luminosity and therefore distances along the $y$-axis are no longer linear.

Some of the previously published lightcurves include symmetric errorbars, while still plotting the logarithm of the luminosity.
This is commonly accomplished by plotting the \emph{relative} error,
given by \citep[see, for example,][]{baird95}
\begin{equation}
  \delta{z} = \delta[\log(y)] \approx d[\log(y)] = \frac{1}{\ln(10)} \frac{dy}{y} \approx 0.434 \frac{\delta y}{y}.
\end{equation}
However, the approximations made to treat the error as a differential only holds for small errors.
The uncertainties in the bolometric luminosities calculated by \texttt{SuperBoL} are not small, and so we choose to plot the absolute errors.
A full table of our calculated bolometric luminosities is given in
\autoref{tab:lightcurves}

The bolometric lightcurves of SN~1998A are shown in \autoref{fig:98a_lc}. The quasi-bolometric luminosities published by \cite{pastorello_sn_2005} are included, and are marginally higher than the quasi-bolometric results of \texttt{SuperBoL}, but lie within the uncertainties.
Both the direct integration and bolometric correction lightcurves produced by \texttt{SuperBoL} are brighter than the direct integration lightcurve of SN~1987A published in \cite{bersten_bolometric_2009}.
The uncertainties in the direct integration lightcurve of SN~1998A are $\delta L_D \approx 0.5L_D$ while the uncertainties in the bolometric correction lightcurve are $\delta L_{BC} \approx 0.4 L_{BC}$.
The large uncertainties in the luminosities are due mainly to the uncertainty in the published distance to SN~1998A \citep{pastorello_sn_2005} shown in \autoref{tab:sn_parameters}.

In the case of SN~2000cb, there is more available photometry in the literature, taken from \citet{hamuy_type_2001} and \citet{kleiser_peculiar_2011}. There is also a previously published lightcurve using an early version of the bolometric correction technique, published in \citet{hamuy_type_2001}.
Our results are similar to those of SN~1998A, with the quasi-bolometric luminosity lower than the direct integration and bolometric correction lightcurves, which are in close agreement with one another.
The lightcurves of SN~2000cb are shown in \autoref{fig:00cb_lc}.
In this supernova, we see that the older version of the bolometric correction technique produces luminosities very similar to those derived from the \texttt{SuperBoL} routine.
The uncertainties are also large, owing to the uncertainty in supernova distance as published in \citet{kleiser_peculiar_2011}.
The direct integration technique has uncertainties of $\delta L_D \approx 0.5 L_D$, and the bolometric correction technique has uncertainties of $\delta L_{BC} \approx 0.3 L_{BC}$.

Our lightcurves for Supernovae SN~2006V and SN~2006au are shown in \autoref{fig:06v_lc} and \autoref{fig:06au_lc}, respectively.
The closest match to previously published results from \citet{taddia_type_2012} is made by the bolometric correction technique, as the direct integration technique produces luminosities systematically higher for both supernovae.
The primary difference between the direct integration technique used by \texttt{SuperBoL} and the method used by \citet{taddia_type_2012} is the interpolation scheme used to integrate between the observed fluxes.
\texttt{SuperBoL} uses a linear interpolation, while the previously published lightcurve uses a cubic spline interpolation as discussed in \autoref{subsec:fqbol}.

Immediately apparent when comparing our lightcurves of the two supernovae is the apparent ``noise'' in the lightcurve of SN~2006au.
This is because of missing observations in the photometric dataset published by \citet{taddia_type_2012}.
\texttt{SuperBoL} currently makes no attempt to interpolate missing magnitudes in a datatset using observed magnitudes from previous or subsequent nights.
As a result, the quasi-bolometric flux on a night with a missing bandpass will be lower than the flux on nights with more complete data.
This has an effect on both the quasi-bolometric and the direct integration luminosities.

For both of these supernovae, the bolometric correction lightcurve is calculated using only $B - V$, since the Carnegie Supernova Project uses the filterset $ugriBV$, and the bolometric correction method as implemented in \texttt{SuperBoL} relies on $B - V$, $V - I$ and $B - I$ colors.
Because of the relatively small uncertainties in the published distances to these supernovae given in \cite{taddia_type_2012}, the errorbars on our calculated lightcurves for SN~2006V and SN~2006au are among the smallest in our sample.
For SN~2006V, the direct integration technique has uncertainties of $\delta L_D \approx 0.14 L_D$, and the bolometric correction technique has uncertainties of $\delta L_{BC} \approx 0.17 L_{BC}$.
For SN~2006au, the direct integration technique has uncertainties equal to those of SN~2006V, with bolometric correction technique uncertainties of $\delta L_{BC} \approx 0.2 L_{BC}$.

The lightcurves of SN~2009E calculated with \texttt{SuperBoL} are shown along with the previously published lightcurve from \citet{pastorello_sn_2012} in \autoref{fig:09e_lc}.
This object was extensively monitored by amateur astronomers during its evolution, resulting in a lightcurve very well-sampled in time \citep[see][\S2.3]{pastorello_sn_2012}.
Most of the images were captured unfiltered --- so although there is abundant photometry available, the limited use of bandpasses means that the number of days on which enough filters were available to re-construct a quasi-bolometric flux are few.
The small uncertainties in the lightcurves ($\delta L_{D} \approx 0.19 L_D$ and $\delta L_{BC} \approx 0.2 L_{BC}$) are again due to the small published uncertainty in the distance to the object.

\subsection{Nickel Mass Estimates}

The post-plateau luminosity of a Type II~P supernova comes primarily from energy deposited by the gamma-rays produced in the decay chain $^{56}$Ni $\rightarrow$ $^{56}$Co $\rightarrow$ $^{56}$Fe.
We use the $\gamma$-ray specific energy for pure $^{56}$Ni from \cite{sutherland_models_1984}:

\begin{equation}
    s = 3.90 \times 10^{10} e^{-\gamma_1 t} + 6.78 \times 10^{9} \left( e^{-\gamma_2 t} - e^{\gamma_1 t} \right)
\end{equation}
where $s$ is in erg s$^{-1}$ g$^{-1}$. The constants $\gamma_1 = \unit[1.32\times10^{-6}]{s^{-1}}$, and $\gamma_2 = \unit[1.02\times10^{-7}]{s^{-1}}$ are the decay rates of $^{56}$Ni and $^{56}$Co, corresponding to half-lives of \unit[6.08]{d} and \unit[78.65]{d}, respectively.

The nickel mass ejected by the supernova can be estimated by fitting the luminosity of the post-plateau tail with the equation

\begin{equation}
    L_{\text{Ni}} = sM_{\text{Ni}}
\end{equation}

\begin{figure}
\centering
    \includegraphics[scale=0.6]{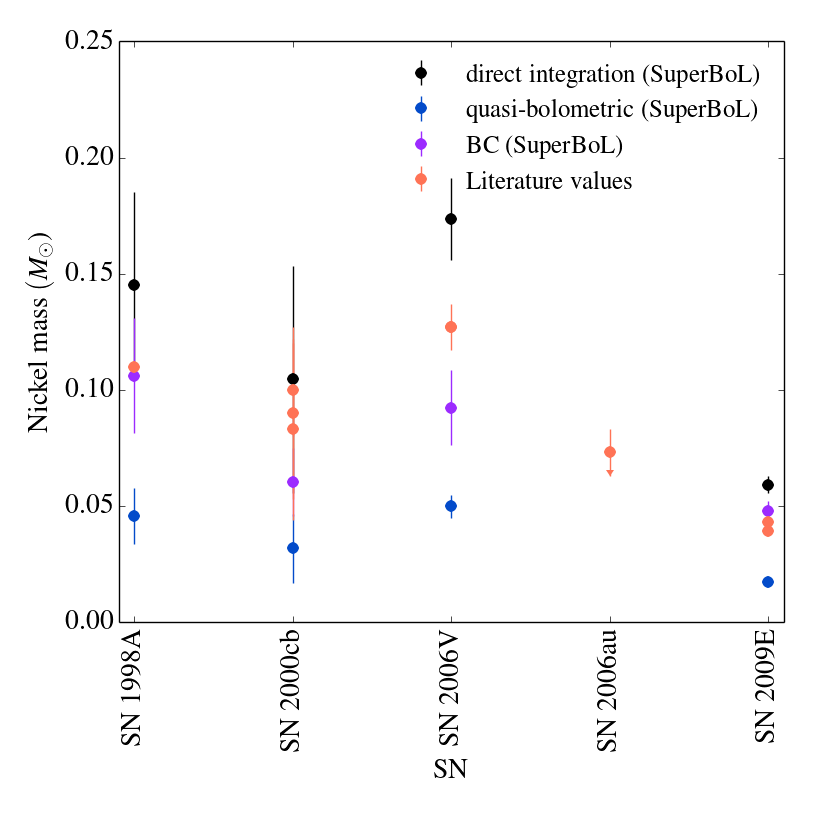}
    \label{fig:ni_masses}
        \caption{Comparison of ejected $^{56}$Ni masses generated by \texttt{SuperBoL} and those previously published in the literature.}
\end{figure}

In our calculations, we again use the \texttt{curve\_fit} function from \texttt{scipy} to determine the mass of $^{56}$Ni ejected by each supernova.
We included bolometric lightcurve points between 120 and 350 days post-explosion in our fit.
The results of our best-fit $^{56}$Ni masses are shown in \autoref{tab:ni_mass}, with one standard deviation errors determined by calculating the square root of the variance $\sigma_{X}^2$ output by \texttt{curve\_fit}.
Also shown in \autoref{tab:ni_mass} are values for the ejected mass of $^{56}$Ni from previously published literature.
The results are also shown in \autoref{fig:ni_masses}, where the horizontal axis serves to separate the results from the different supernovae.
Excluded from our analysis is SN~2006au, which does not include any observations which conclusively fall on the radioactive tail.

The differences between the lightcurves calculated by \texttt{SuperBoL} and those previously published in the literature are reflected here in the inferred nickel masses.
Where the direct integration technique produced luminosities significantly greater than those in previously published studies (most notably in the cases of SN~1998A and SN~2006V,) the resulting nickel mass was correspondingly higher.
Similarly, it is clear that using quasi-bolometric luminosities will result in nickel masses many times smaller than the other techniques analyzed here.

It is encouraging to note that our results for the mass of $^{56}$Ni using the direct integration and bolometric correction techniques broadly agree with those produced by semi-analytic lightcurve models and hydrodynamic models used previously in the literature.
In the case of SN~1998A, \citet{pastorello_sn_2005} used the semi-analytic code developed by \citet{zampieri_peculiar_2003} to produce a best-fit bolometric lightcurve.
The $^{56}$Ni mass ejected by the supernova is one of the input parameters of the model, along with other progenitor and explosion parameters, outlined in \citet{pastorello_sn_2005} (their Table 5.)
Other studies have used the hydrodynamic models of \citet{young_parameter_2004} \citep[][in the case of SN~2000cb]{kleiser_peculiar_2011}, the semi-analytic models of \citet{imshennik_construction_1992} \citep[][in the case of SN~2006V and and SN~2006au]{taddia_type_2012}, and the hydrodynamic models of \citet{pumo_numerical_2010} and \citet{pumo_radiation-hydrodynamical_2011} \citep[][in the case of SN~2009E]{pastorello_sn_2012}

\subsection{Peak Luminosities}

Another observationally meaningful parameter is the peak luminosity of a supernova, $L_{\text{peak}}$.
We measure $L_{\text{peak}}$ as the highest of our bolometric luminosities, and compare that with values taken from the literature in \autoref{tab:sn_Lpeak}.
The first three columns include the uncertainties output by \texttt{SuperBoL}, and the last column includes, where possible, uncertainties reported in the literature.
We note that the uncertainties in our values for the peak bolometric luminosity are very large compared to those previously published in the literature.
The cause of this discrepancy is unclear.
Not all published bolometric lightcurves include errorbars, and those that do often lack a detailed description of the uncertainty calculations.

As mentioned in \autoref{sec:sn_parameters}, the meaningful comparisons in \autoref{tab:sn_Lpeak} are between the different methods used for calculating the bolometric luminosity of a single supernova.
Inter-comparisons between supernovae are hampered by the different methods of making distance estimates, and the different cosmological parameters used in the redshift-distance relations by different groups.

\subsection{Nickel - Luminosity Relation}

\begin{figure}
\centering
    \includegraphics[scale=0.6]{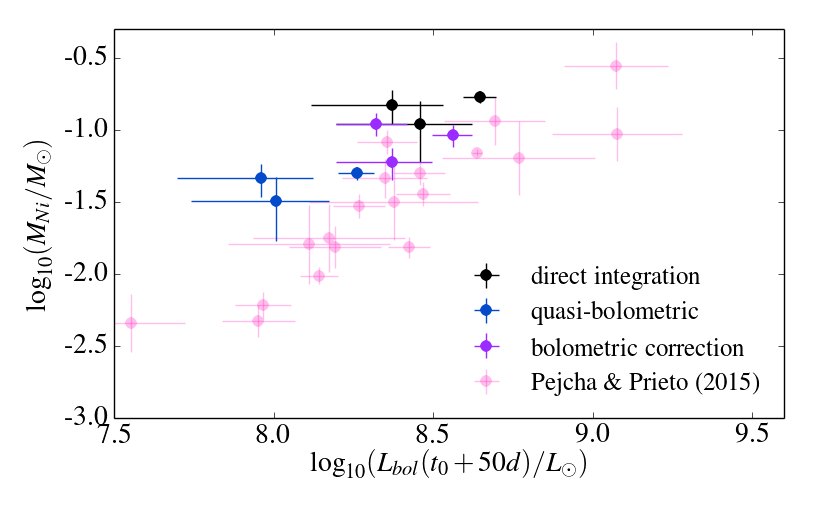}
    \label{fig:ni_l50}
    \caption{Relation between the ejected nickel mass and the bolometric luminosity of the supernova measured at $t_0 + 50$d. SN~2009E is excluded from this comparison, because no photometry is available around $t_0 + 50$d}
\end{figure}

The radioactive decay of $^{56}$Ni is known to power the lightcurves of supernovae during late stages in their evolution.
For Type~Ia supernovae, the lightcurve is entirely driven by radioactive decay, which leads to a natural relationship between the peak luminosity of the supernova and the mass of ejected $^{56}$Ni known as ``Arnett's Law'' \citep{arnett_type_1982, arnett_hubbles_1985, branch_hubble_1992}. For Type~II-P supernovae, there is also a correlation between plateau luminosity and ejected $^{56}$Ni mass \citep{hamuy_observed_2003, spiro_low_2014}.
In \autoref{fig:ni_l50}, we add the supernovae in our sample to the dataset published by \citet{pejcha_global_2015} (their figure 15, after removal of the outlier SN~2007od for clarity.)

It should be noted that in producing these plots, we are violating the principle laid out in \autoref{sec:sn_parameters}, and inter-comparing the luminosities and ejected nickel masses of the supernovae in our sample.
These results, therefore, should be viewed as preliminary.

While the three different methods used in this paper for calculating the bolometric luminosity produce results that follow a similar trend to the previously published results, the luminosities of the peculiar Type~II-P supernovae in this sample appear systematically lower for a given value of the nickel mass than their counterparts in \citet{pejcha_global_2015}.
This is likely due to the slow rise times which make this set of supernovae photometrically distinct.
The choice of measuring the plateau luminosity at $t_0 + 50$d in \citet{pejcha_global_2015} works for typical Type~II-P supernovae, with rise times of $\sim 10$ days.
The SN~1987A-like supernovae in this sample, however, rise to peak luminosity much more slowly, and so are systematically dimmer at $t_0 + 50$d.

\begin{figure}
\centering
    \includegraphics[scale=0.6]{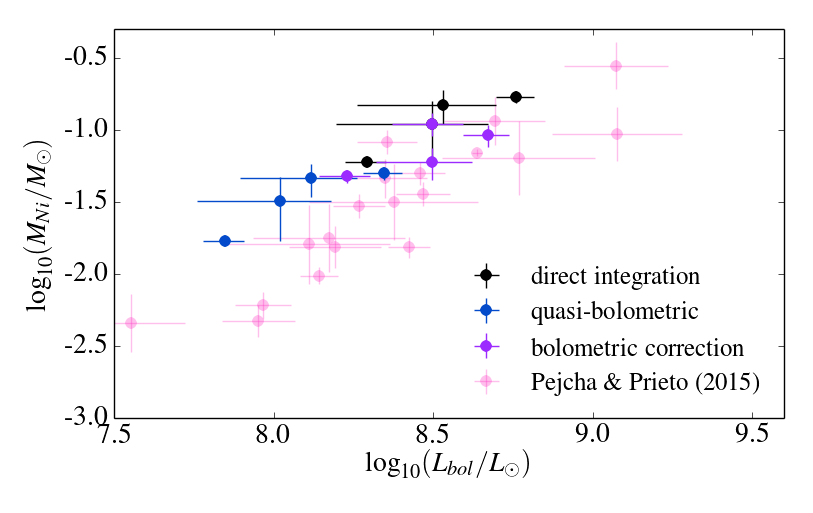}
    \label{fig:ni_lpeak}
    \caption{Relation between the ejected nickel mass and the bolometric luminosity of the supernova measured at peak brightness. The data from \citet{pejcha_global_2015} are still measured at $t_0 + 50$d.}
\end{figure}

To somewhat remedy this, we instead compare the peak luminosity of our supernovae to the same dataset. This is shown in \autoref{fig:ni_lpeak}.
The SN~1987A-like supernovae are still dimmer for a given value of the ejected nickel mass, but are shifted closer to the main trend.

Another possible interpretation of the data is that these SN~1987A-like supernovae are producing more radioactive $^{56}$Ni than more typical Type~II-P supernovae at a given plateau luminosity.
If the progenitors of these SN~1987A-like supernovae are blue supergiant stars as was the progenitor of SN~1987A and as has been suggested by past studies of these supernovae, this may be an observable signature of the the progenitor star or explosion characteristics that set these supernovae apart from the general population of Type~II-P supernovae.

\section{Discussion}{\label{sec:discussion}}

The results in \autoref{sec:results} show that the different methods used in the literature to calculate the bolometric luminosity of a supernova can differ significantly.

The quasi-bolometric technique seems to be useful only in making comparisons between two supernovae with identical wavelength coverage.
As shown in the results from SN~2006au in \autoref{sec:lightcurves}, the inclusion or exclusion of a bandpass can result in large variations in the resulting bolometric luminosity.
The quasi-bolometric technique will produce results several times lower than those of the bolometric correction or direct integration methods.

The direct integration and bolometric correction schemes were in rough agreement for these five supernovae.
It should be noted that the two supernovae with the largest difference between these two methods are SN~2006V and SN~2006au --- supernovae for which only the $B-V$ color could be used in the bolometric correction. In the other supernovae, the results from  using all available combinations of $B-V$, $V-I$, and $B-I$ were averaged together to produce $L_{\text{BC}}$.
It seems possible that this averaging resulted in the closer agreement between the bolometric correction results and the direct integration results.

It is apparent from our calculations that the shape of the bolometric lightcurve of SN~2000cb stands out among this sample of peculiar Type II-P supernovae.
The rise time to near-maximum light is faster and the plateau-phase is broader than the other supernovae in the sample.
\citet{utrobin_supernova_2011} hypothesized that this difference may be due to mixing of $^{56}$Ni out to ejecta velocities of \unit[8400]{km} s$^{-1}$, much farther than the modest mixing out to \unit[3000]{km} s$^{-1}$ assumed for SN~1987A.
Whatever the cause of the relatively fast rise, it is an effect that is clearly not an artifact of the technique used to calculate the bolometric luminosity --- the slope of the early lightcurve is steeper than the other supernovae regardless of the method used.

\subsection{Comparison with PHOENIX models}

Though we have shown that the direct integration and bolometric correction lightcurves broadly agree for this sample of supernovae, it is impossible to determine if that agreement reflects the true bolometric luminosity of the supernova.
To make progress toward verifying the results of the direct integration and bolometric correction techniques, we would need an independent method of determining the bolometric luminosity of the supernova --- one that did not depend on observed photometry like all the techniques studied here.

To that end, we plan to make use of the NLTE radiative transfer code
\texttt{PHOENIX} \citep[see, for example,][]{hauschildt_numerical_1999}.
\texttt{PHOENIX} calculates the full spectral energy distribution of a supernova, allowing for the use of synthetic photometry to determine the absolute magnitudes of the SN in various photometric systems.
In a future study, we will use \texttt{PHOENIX} to fit the observed spectra of the supernovae studied here, then feed the synthetic photometry of those computed spectra through \texttt{SuperBoL} to determine how closely each technique reproduces the bolometric luminosity reported by \texttt{PHOENIX}.
This will also give us a way of better estimating any systematic errors in the luminosity calculation techniques.

As a proof-of-concept, we fed the synthetic photometry of an unrelated \texttt{PHOENIX} spectrum through \texttt{SuperBoL}.
In an attempt to avoid biasing the results of this test, the author running \texttt{SuperBoL} (JL) was only given the synthetic photometry --- the luminosity of the underlying \texttt{PHOENIX} model was not revealed until after the results had been calculated. 

Our initial calculations are promising. The bolometric luminosity of the synthetic spectrum was $L_{\text{bol}} = \unit[11.08\times10^{41}]{erg}$ s$^{-1}$.
The quasi-bolometric technique, using the full compliment of bandpasses produced a luminosity of $L_{\text{qbol}} = 7.25\times10^{41}$ erg s$^{-1}$.
The direct integration routine calculated a bolometric luminosity of $L_{\text{D}} = \unit[11.30\times 10^{41}]{erg}$ s$^{-1}$ when utilizing all available bandpasses.
The bolometric correction method resulted in $L_{\text{BC}} = \unit[8.61\times10^{41}]{erg}$ s$^{-1}$ when averaging the results of $B - V$, $V - I$, and $B - I$ calculations.
The value calculated using the direct integration technique is remarkably close (within 5\%) of the true luminosity of the synthetic spectrum.
The luminosity determined through the bolometric correction technique is lower, but within 25\% of the true luminosity.

In order to check the dependence of our methods on the completeness of the photometric observations, we re-calculated the bolometric luminosity of the \texttt{PHOENIX} spectrum using several smaller subsets of the available synthetic magnitudes.
The results are shown in \autoref{tab:sparse_phot}.

As was evident in the bolometric lightcurves of SN~2006au in \autoref{sec:lightcurves}, the quasi-bolometric technique is very sensitive to the wavelength range of the observed photometry.
Since no corrections are made for flux which falls outside the observed range, this method consistently under-estimates the bolometric luminosity of the \texttt{PHOENIX} model by a significant amount.

One promising result is the consistency in the luminosity calculated using the direct integration technique, even when the wavelength range of the observed photometry varies significantly.
Based on the design of the technique, this is to be expected --- since missing flux blueward and redward of the observed wavelength range is filled in using the assumption of a blackbody SED.
The concordance between the expected behavior of the technique and the results produced by \texttt{SuperBoL}, and the close match between the calculated and model luminosity for all subsets of the observed photometry, are encouraging.

The results obtained by the bolometric correction technique are more puzzling.
The luminosity shows a large variation, depending on which of the three filter combinations are available for averaging.
Whether this is inherent in the technique, or a result of peculiarities with the underlying \texttt{PHOENIX} model will be one of the questions addressed in our future work.
We hope that by using new \texttt{PHOENIX} models matched to each of the supernovae in our current study, we can isolate the cause of this variation.
We will also include the bolometric correction technique used in \citet{pejcha_global_2015} and examine whether this variability is common to the entire class of bolometric correction methods.

\section{Conclusions}

We have constructed the publicly available code \texttt{SuperBoL} to produce bolometric
lightcurves from broad-band photometry and applied it to Type II supernovae
thought to have blue supergiant progenitors. We have shown that the results from \texttt{SuperBoL} agree reasonably well
when compared to techniques used previously in the literature, although some
discrepancies exist. In future work, we will compare our results with the bolometric luminosities of
\texttt{PHOENIX} models for the compact progenitor case. Our code should be
especially useful for exploring not-well characterized supernovae,  in
particular, extremely luminous supernovae (ELSNe) for both type ELSNe I/II and
fast dim objects where the progenitor may or may not be a core
collapse. It is our hope that our code will be improved upon in the future through contributions from the supernova community, and we welcome suggestions for improvements or new features.

\section{Acknowledgments}

This work made use of Astropy, a community-developed core Python package for Astronomy \citep{astropy_collaboration_astropy:_2013}.
The work has been supported in part by
support for programs
HST-GO-12948.004-A was provided by NASA through a grant from the
Space Telescope Science Institute, which is operated by the
Association of Universities for Research in Astronomy, Incorporated,
under NASA contract NAS5-26555.
This work was also supported in part by 
NSF grant AST-0707704, by NASA Grant NNX16AB25G and DOE Grant DE-SC0009956.

This research used resources of the
National Energy Research Scientific Computing Center (NERSC), which is
supported by the Office of Science of the U.S.  Department of Energy
under Contract No.  DE-AC02-05CH11231; and the H\"ochstleistungs
Rechenzentrum Nord (HLRN).  We thank both these institutions for a
generous allocation of computer time. 

\clearpage

\bibliography{bsg_lbol}

\pagebreak

\capstartfalse
\begin{deluxetable}{ccccc}
    \tabletypesize{\footnotesize}
    \tablecolumns{4}
    \tablewidth{0pt}
    \tablecaption{Adopted SN parameters from the literature \label{tab:sn_parameters}}
    \tablehead{
    \colhead{SN} & \colhead{$D$ (Mpc)} & \colhead{Explosion (JD)} & \colhead{$A_V$(total)} & \colhead{Ref}}
    \startdata
    1998A & $30\pm7$ &  $2450801\pm4$ & 0.399 & \tablenotemark{a}\\
    2000cb & $30\pm7$ & $2451656\pm4$ & 0.373 & \tablenotemark{b}\\
    2006V & $73\pm5$ & $2453748\pm4$ & 0.09 & \tablenotemark{c}\\
    2006au & $46\pm3$ & $2453794\pm9$ & 0.97 & \tablenotemark{c}\\
    2009E & $30\pm2$ & $2454833\pm3$ & 0.124 & \tablenotemark{d}
    \enddata
    \tablerefs{\textsuperscript{a}\cite{pastorello_sn_2005}, \textsuperscript{b}\cite{kleiser_peculiar_2011}, \textsuperscript{c}\cite{taddia_type_2012}, \textsuperscript{d}\cite{pastorello_sn_2012}}
\end{deluxetable}
\capstarttrue

\capstartfalse
\begin{deluxetable}{ccccccc}
    \tabletypesize{\footnotesize}
    \tablecolumns{7}
    \tablewidth{0pt}
    \tablecaption{Bolometric Lightcurves \label{tab:lightcurves}}
    \tablehead{
    \colhead{SN} & \colhead{JD} & \colhead{Phase} & \colhead{$L_{\text{qbol}}$} & \colhead{$L_{\text{D}}$} & \colhead{$L_{\text{BC}}$} & \colhead{$L_\text{lit}$} \\
                 & & \colhead{(days)} & \colhead{10$^{41}$ (erg/s)} & \colhead{10$^{41}$ (erg/s)} & \colhead{10$^{41}$ (erg/s)} & \colhead{10$^{41}$ (erg/s)}}
    \startdata
    SN~1998A  & 2450837.8 & $37 \pm 4$  & $2.9 \pm 1.3$ & $7 \pm 3$  & $6.2 \pm 1.7$   & $3.6 \pm 0.4$\tablenotemark{a}\\
              & 2450845.9 & $45 \pm 4$  & \nodata       & \nodata    & $8 \pm 4$       & $4.2 \pm 0.4$\tablenotemark{a}\\
              & 2450846.7 & $46 \pm 4$  & $3.5 \pm 1.6$ & $9 \pm 4$  & $8 \pm 2$       & $4.4 \pm 0.8$\tablenotemark{a}\\
              & \nodata   & $50 \pm 4$  & \nodata       & \nodata    & \nodata         & $4.5 \pm 0.6$\tablenotemark{a}\\
              & 2450898.5 & $98 \pm 4$  & $5 \pm 2$     & $13 \pm 6$ & $12 \pm 3$      & $6.3 \pm 1.0$\tablenotemark{a}\\
              & 2450899.8 & $99 \pm 4$  & \nodata       & \nodata    & $11 \pm 5$      & $6.0 \pm 1.2$\tablenotemark{a}\\
              & 2450939.6 & $139 \pm 4$ & $1.7 \pm 0.8$ & $5 \pm 3$  & $5 \pm 2$       & $2.1 \pm 0.7$\tablenotemark{a}\\
              & 2450960.5 & $160 \pm 4$ & \nodata       & \nodata    & \nodata         & $1.9 \pm 0.4$\tablenotemark{a}\\
              & 2450962.5 & $162 \pm 4$ & $1.6 \pm 0.7$ & $4 \pm 2$  & $3.3 \pm 1.2$   & $1.9 \pm 0.3$\tablenotemark{a}\\
              & 2450991.5 & $190 \pm 4$ & $1.1 \pm 0.5$ & $5 \pm 2$  & $2.7 \pm 1.1$   & $1.4 \pm 0.4$\tablenotemark{a}\\
              & 2451143.8 & $344 \pm 4$ & \nodata       & \nodata    & \nodata         & $0.35 \pm 0.10$\tablenotemark{a}\\
              & 2451200.7 & $400 \pm 4$ & \nodata       & \nodata    & $0.23 \pm 0.11$ & $0.19 \pm 0.09$\tablenotemark{a}\\
    \hline\\
\tablebreak
    SN~2000cb & 2451663.81 & $8 \pm 4$   & $1.1 \pm 0.5$ & $2.1 \pm 0.9$ & $2.0 \pm 0.6$   &$2.5$\tablenotemark{b}\\
              & 2451663.92 & $8 \pm 4$   & \nodata       & \nodata       & $2.4 \pm 0.7$   &\nodata\tablenotemark{c}\\
              & 2451667.86 & $12 \pm 4$  & \nodata       & \nodata       & $2.6 \pm 0.7$   &\nodata\tablenotemark{c}\\
              & 2451675.70 & $20 \pm 4$  & $2.1 \pm 1.0$ & $5 \pm 2$     & $4.5 \pm 1.2$   &$5.2$\tablenotemark{b}\\
              & 2451676.76 & $21 \pm 4$  & $2.3 \pm 1.1$ & $5 \pm 3$     & $4.8 \pm 1.3$   &\nodata\tablenotemark{b}\\
              & 2451677.77 & $22 \pm 4$  & $2.4 \pm 1.1$ & $6 \pm 3$     & $5.1 \pm 1.4$   &$5.9$\tablenotemark{b}\\
              & 2451681.74 & $26 \pm 4$  & $2.8 \pm 1.3$ & $7 \pm 3$     & $6.0 \pm 1.6$   &\nodata\tablenotemark{b}\\
              & 2451682.81 & $27 \pm 4$  & $3.0 \pm 1.4$ & $7 \pm 3$     & $6.2 \pm 1.7$   &$7.0$\tablenotemark{b}\\
              & 2451683.75 & $28 \pm 4$  & \nodata       & \nodata       & $6 \pm 3$       &\nodata\tablenotemark{b}\\
              & 2451683.78 & $28 \pm 4$  & $3.1 \pm 1.4$ & $7 \pm 3$     & $6.4 \pm 1.8$   &\nodata\tablenotemark{b}\\
              & 2451684.75 & $29 \pm 4$  & $3.3 \pm 1.5$ & $7 \pm 3$     & $7 \pm 3$       &$7.7$\tablenotemark{b}\\
              & 2451692.88 & $37 \pm 4$  & \nodata       & \nodata       & $8 \pm 2$       &\nodata\tablenotemark{c}\\
              & 2451695.57 & $40 \pm 4$  & $3.7 \pm 1.7$ & $9 \pm 4$     & $8 \pm 2$       &$9.0$\tablenotemark{b}\\
              & 2451696.87 & $44 \pm 4$  & \nodata       & \nodata       & $8 \pm 2$       &\nodata\tablenotemark{c}\\
              & 2451699.72 & $44 \pm 4$  & $3.8 \pm 1.8$ & $10 \pm 5$    & $9 \pm 2$       &$9.3$\tablenotemark{b}\\
              & 2451700.85 & $45 \pm 4$  & \nodata       & \nodata       & $9 \pm 2$       &\nodata\tablenotemark{c}\\
              & 2451705.70 & $50 \pm 4$  & $3.8 \pm 1.8$ & $10 \pm 5$    & $9 \pm 2$       &$9.3$\tablenotemark{b}\\
              & 2451706.85 & $51 \pm 4$  & \nodata       & \nodata       & $9 \pm 2$       &\nodata\tablenotemark{c}\\
              & 2451713.82 & $58 \pm 4$  & \nodata       & \nodata       & $9 \pm 3$       &\nodata\tablenotemark{c}\\
              & 2451717.84 & $62 \pm 4$  & \nodata       & \nodata       & $10 \pm 3$      &\nodata\tablenotemark{c}\\
              & 2451721.82 & $66 \pm 4$  & \nodata       & \nodata       & $11 \pm 3$      &\nodata\tablenotemark{c}\\
              & 2451728.77 & $73 \pm 4$  & \nodata       & \nodata       & $11 \pm 3$      &\nodata\tablenotemark{c}\\
              & 2451730.67 & $75 \pm 4$  & $3.9 \pm 1.8$ & $12 \pm 6$    & $11 \pm 3$      &$10.3$\tablenotemark{b}\\
              & 2451735.72 & $80 \pm 4$  & \nodata       & \nodata       & $10 \pm 3$      &\nodata\tablenotemark{c}\\
              & 2451738.64 & $83 \pm 4$  & $3.8 \pm 1.8$ & $12 \pm 5$    & $12 \pm 3$      &$10.6$\tablenotemark{b}\\
              & 2451742.73 & $87 \pm 4$  & \nodata       & \nodata       & $12 \pm 4$      &\nodata\tablenotemark{c}\\
              & 2451745.66 & $90 \pm 4$  & $3.5 \pm 1.6$ & $11 \pm 5$    & $11 \pm 3$      &$9.7$\tablenotemark{b}\\
              & 2451749.69 & $94 \pm 4$  & \nodata       & \nodata       & $11 \pm 3$      &\nodata\tablenotemark{c}\\
              & 2451752.70 & $97 \pm 4$  & \nodata       & \nodata       & $9 \pm 5$       &\nodata\tablenotemark{c}\\
              & 2451756.69 & $101 \pm 4$ & \nodata       & \nodata       & $8 \pm 3$       &\nodata\tablenotemark{c}\\
              & 2451757.64 & $102 \pm 4$ & $2.5 \pm 1.2$ & $9 \pm 4$     & $10 \pm 3$      &$7.7$\tablenotemark{b}\\
              & 2451781.66 & $126 \pm 4$ & \nodata       & \nodata       & $3.9 \pm 1.6$   &\nodata\tablenotemark{c}\\
              & 2451788.66 & $133 \pm 4$ & \nodata       & \nodata       & $1.9 \pm 1.0$   &\nodata\tablenotemark{c}\\
              & 2451795.49 & $140 \pm 4$ & $1.2 \pm 0.6$ & $4.1 \pm 1.9$ & $12 \pm 7$      &$3.7$\tablenotemark{b}\\
              & 2451795.64 & $140 \pm 4$ & \nodata       & \nodata       & $2.3 \pm 0.9$   &\nodata\tablenotemark{c}\\
    \hline\\
\tablebreak
    SN~2006V  & 2453773.71 & $26 \pm 4$  & $3.8 \pm 0.5$ & $8.9 \pm 1.2$ & $7.1 \pm 1.2$   & $7.5 \pm 0.3$\tablenotemark{d}\\
              & 2453774.82 & $27 \pm 4$  & $4.0 \pm 0.5$ & $9.3 \pm 1.3$ & $7.4 \pm 1.3$   & $7.8 \pm 0.3$\tablenotemark{d}\\
              & 2453775.67 & $28 \pm 4$  & $4.1 \pm 0.6$ & $9.6 \pm 1.3$ & $7.7 \pm 1.3$   & $8.1 \pm 0.3$\tablenotemark{d}\\
              & 2453778.81 & $31 \pm 4$  & $4.6 \pm 0.6$ & $10.6 \pm 1.5$& $8.7 \pm 1.5$   & $9.0 \pm 0.3$\tablenotemark{d}\\
              & 2453784.90 & $37 \pm 4$  & \nodata       & \nodata       & $10.2 \pm 1.8$  & \nodata\tablenotemark{d} \\
              & 2453786.88 & $39 \pm 4$  & $5.8 \pm 0.8$ & $13.1 \pm 1.8$& $10.8 \pm 1.9$  & $10.9 \pm 0.2$\tablenotemark{d}\\
              & 2453795.84 & $48 \pm 4$  & $6.2 \pm 0.8$ & $16 \pm 2$    & $13 \pm 2$      & $12.9 \pm 0.3$\tablenotemark{d}\\
              & 2453799.75 & $52 \pm 4$  & $6.7 \pm 0.9$ & $17 \pm 2$    & $14 \pm 2$      & $13.8 \pm 0.4$\tablenotemark{d}\\
              & 2453804.84 & $57 \pm 4$  & $7.2 \pm 1.0$ & $19 \pm 3$    & $15 \pm 3$      & \nodata\tablenotemark{d}\\
              & 2453805.79 & $58 \pm 4$  & $7.3 \pm 1.0$ & $19 \pm 3$    & $15 \pm 3$      & $15.3 \pm 0.4$\tablenotemark{d}\\
              & 2453818.77 & $71 \pm 4$  & $8.3 \pm 1.1$ & $22 \pm 3$    & $17 \pm 3$      & $17.7 \pm 0.4$\tablenotemark{d}\\
              & 2453824.73 & $77 \pm 4$  & $8.5 \pm 1.2$ & $22 \pm 3$    & $18 \pm 3$      & $18.2 \pm 0.4$\tablenotemark{d}\\
              & 2453832.75 & $85 \pm 4$  & $8.2 \pm 1.1$ & $22 \pm 3$    & $17 \pm 3$      & $17.7 \pm 0.4$\tablenotemark{d}\\
              & 2453838.75 & $91 \pm 4$  & $7.6 \pm 1.0$ & $20 \pm 3$    & $16 \pm 3$      & $16.4 \pm 0.4$\tablenotemark{d}\\
              & 2453846.73 & $99 \pm 4$  & $6.0 \pm 0.8$ & $17 \pm 2$    & $12 \pm 2$      & $13.4 \pm 0.3$\tablenotemark{d}\\
              & 2453853.64 & $106 \pm 4$ & $4.1 \pm 0.6$ & $12.5 \pm 1.7$& $8.2 \pm 1.4$   & $10.0 \pm 0.3$\tablenotemark{d}\\
              & 2453862.57 & $115 \pm 4$ & $2.8 \pm 0.4$ & $9.0 \pm 1.2$ & $5.4 \pm 0.9$   & $7.2 \pm 0.3$\tablenotemark{d}\\
              & 2453867.57 & $120 \pm 4$ & $2.6 \pm 0.4$ & $8.0 \pm 1.1$ & $5.5 \pm 1.1$   & $6.5 \pm 0.3$\tablenotemark{d}\\
              & 2453892.56 & $145 \pm 4$ & $1.8 \pm 0.3$ & $6.4 \pm 0.9$ & \nodata         & $5.0 \pm 0.3$\tablenotemark{d}\\
              & 2453898.56 & $151 \pm 4$ & $1.8 \pm 0.3$ & $6.4 \pm 0.9$ & $3.3 \pm 0.6$   & $4.5 \pm 0.2$\tablenotemark{d}\\
    \hline\\
\tablebreak
    SN~2006au & 2453805.89 & $12 \pm 9$  & $5.5 \pm 0.8$ & $13.9 \pm 1.9$& $9.2 \pm 1.6$   & $10.4 \pm 0.5$\tablenotemark{d}\\
              & 2453809.85 & $16 \pm 9$  & $4.7 \pm 0.7$ & $12.5 \pm 1.7$& $9.6 \pm 1.7$   & $9.6 \pm 0.4$\tablenotemark{d}\\
              & 2453815.86 & $22 \pm 9$  & $4.9 \pm 0.7$ & $12.6 \pm 1.8$& $8.7 \pm 1.5$   & $9.2 \pm 0.4$\tablenotemark{d}\\
              & 2453818.90 & $25 \pm 9$  & $5.0 \pm 0.7$ & $12.2 \pm 1.7$& $8.5 \pm 1.5$   & $9.3 \pm 0.3$\tablenotemark{d}\\
              & 2453819.90 & $26 \pm 9$  & $3.3 \pm 0.5$ & $15 \pm 2$    & $8.7 \pm 1.5$   & $9.4 \pm 0.7$\tablenotemark{d}\\
              & 2453823.83 & $30 \pm 9$  & $5.2 \pm 0.7$ & $12.0 \pm 1.7$& $8.6 \pm 1.5$   & $9.4 \pm 0.4$\tablenotemark{d}\\
              & 2453824.87 & $31 \pm 9$  & $5.3 \pm 0.7$ & $12.3 \pm 1.7$& $8.9 \pm 1.5$   & $9.5 \pm 0.4$\tablenotemark{d}\\
              & 2453828.89 & $35 \pm 9$  & $2.4 \pm 0.3$ & $16 \pm 2$    & $9.6 \pm 1.6$   & $10.1 \pm 0.8$\tablenotemark{d}\\
              & 2453829.92 & \nodata     & \nodata       & \nodata       & \nodata         & $10.0 \pm 1.2$\tablenotemark{d}\\
              & 2453830.89 & $37 \pm 9$  & $5.5 \pm 0.8$ & $12.7 \pm 1.8$& $8.7 \pm 1.5$   & $9.9 \pm 0.7$\tablenotemark{d}\\
              & 2453831.83 & $38 \pm 9$  & $5.8 \pm 0.8$ & $12.6 \pm 1.8$& \nodata         & $10.1 \pm 0.7$\tablenotemark{d}\\
              & 2453832.86 & $39 \pm 9$  & $2.4 \pm 0.3$ & $14 \pm 2$    & $9.5 \pm 1.6$   & $10.3 \pm 0.8$\tablenotemark{d}\\
              & 2453835.87 & $42 \pm 9$  & $5.9 \pm 0.8$ & $13.2 \pm 1.8$& $9.8 \pm 1.7$   & $10.8 \pm 0.4$\tablenotemark{d}\\
              & 2453838.82 & $45 \pm 9$  & $6.7 \pm 0.9$ & $15 \pm 2$    & $11.2 \pm 1.9$  & $12.0 \pm 0.8$\tablenotemark{d}\\
              & 2453840.83 & $47 \pm 9$  & $6.7 \pm 0.9$ & $15 \pm 2$    & $11 \pm 2$      & $11.9 \pm 0.6$\tablenotemark{d}\\
              & 2453841.85 & $48 \pm 9$  & $6.9 \pm 1.0$ & $15 \pm 2$    & $12 \pm 2$      & $12.1 \pm 0.5$\tablenotemark{d}\\
              & 2453845.82 & $52 \pm 9$  & $6.3 \pm 0.9$ & $17 \pm 2$    & $12 \pm 2$      & $13.0 \pm 0.7$\tablenotemark{d}\\
              & 2453850.81 & $57 \pm 9$  & $6.6 \pm 0.9$ & $18 \pm 2$    & $13 \pm 2$      & $13.6 \pm 0.7$\tablenotemark{d}\\
              & 2453853.85 & $60 \pm 9$  & $6.7 \pm 0.9$ & $18 \pm 2$    & $13 \pm 2$      & $14.0 \pm 0.8$\tablenotemark{d}\\
              & 2453858.79 & $65 \pm 9$  & $7.4 \pm 0.9$ & $21 \pm 3$    & $14 \pm 3$      & $14.9 \pm 0.8$\tablenotemark{d}\\
              & 2453861.82 & $68 \pm 9$  & $7.2 \pm 1.0$ & $20 \pm 3$    & $14 \pm 2$      & $15.2 \pm 0.8$\tablenotemark{d}\\
              & 2453862.80 & $69 \pm 9$  & $4.5 \pm 1.0$ & $22 \pm 3$    & $14 \pm 3$      & $15.6 \pm 1.2$\tablenotemark{d}\\
              & 2453866.75 & $73 \pm 9$  & $7.7 \pm 0.6$ & $20 \pm 3$    & $15 \pm 3$      & $16.0 \pm 0.9$\tablenotemark{d}\\
              & 2453867.78 & $74 \pm 9$  & $7.6 \pm 1.1$ & $21 \pm 3$    & $15 \pm 3$      & $15.7 \pm 1.0$\tablenotemark{d}\\
              & 2453870.80 & $77 \pm 9$  & $7.6 \pm 1.1$ & $21 \pm 3$    & $15 \pm 3$      & $16.2 \pm 0.9$\tablenotemark{d}\\
              & 2453871.79 & $78 \pm 9$  & $7.5 \pm 1.0$ & $21 \pm 3$    & $15 \pm 3$      & $15.8 \pm 0.8$\tablenotemark{d}\\
              & 2453872.78 & $79 \pm 9$  & $7.4 \pm 1.0$ & $21 \pm 3$    & $14 \pm 2$      & $15.7 \pm 1.0$\tablenotemark{d}\\
              & 2453886.78 & $93 \pm 9$  & \nodata       & \nodata       & $8.9 \pm 1.6$   & $11.5 \pm 1.0$\tablenotemark{d}\\
              & 2453890.72 & $97 \pm 9$  & $4.5 \pm 0.6$ & $13.3 \pm 1.9$& $8.2 \pm 1.9$   & $9.7 \pm 0.7$\tablenotemark{d}\\
              & 2453891.74 & $98 \pm 9$  & $4.2 \pm 0.6$ & $14 \pm 2$    & $7.7 \pm 1.3$   & $9.2 \pm 0.5$\tablenotemark{d}\\
              & 2453892.76 & $99 \pm 9$  & $4.0 \pm 0.5$ & $15 \pm 2$    & $7.7 \pm 1.4$   & $9.4 \pm 0.5$\tablenotemark{d}\\
              & 2453893.71 & $100 \pm 9$ & $4.1 \pm 0.7$ & $16 \pm 2$    & $7.2 \pm 1.2$   & $8.5 \pm 0.5$\tablenotemark{d}\\
              & 2453894.75 & $101 \pm 9$ & $3.6 \pm 0.5$ & $11.9 \pm 1.7$& $6.9 \pm 1.2$   & $7.6 \pm 0.4$\tablenotemark{d}\\
              & 2453897.66 & $104 \pm 9$ & \nodata       & \nodata       & $4.5 \pm 0.9$   & $5.3 \pm 0.6$\tablenotemark{d}\\
              & 2453898.71 & $105 \pm 9$ & \nodata       & \nodata       & \nodata         & $4.0 \pm 0.5$\tablenotemark{d}\\
    \hline\\
\tablebreak
    SN~2009E  & 2454840.84 & \nodata     & \nodata       & \nodata       & \nodata         & $1.7$\tablenotemark{e}\\
              & 2454858.01 & $26 \pm 9$  & $1.1 \pm 0.16$ & $2.6 \pm 0.5$ & $2.6 \pm 0.3$  & $2.6$\tablenotemark{e}\\
              & 2454915.41 & $83 \pm 9$  & $2.5 \pm 0.3$  & $6.1 \pm 0.9$ & $5.6 \pm 0.6$  & $6.2$\tablenotemark{e}\\
              & 2454917.33 & $85 \pm 9$  & \nodata        & \nodata       & $6.3 \pm 1.8$  & $6.4$\tablenotemark{e}\\
              & 2454922.49 & $90 \pm 9$  & $2.5 \pm 0.3$  & $6.3 \pm 0.9$ & $5.7 \pm 0.6$  & $6.2$\tablenotemark{e}\\
              & 2454923.34 & $91 \pm 9$  & \nodata        & \nodata       & \nodata        & $6.2$\tablenotemark{e}\\
              & 2454926.42 & $94 \pm 9$  & \nodata        & \nodata       & \nodata        & $6.3$\tablenotemark{e}\\
              & 2454928.33 & $96 \pm 9$  & \nodata        & \nodata       & $6.5 \pm 1.2$  & $6.4$\tablenotemark{e}\\
              & 2454934.39 & $102 \pm 9$ & \nodata        & \nodata       & $6.3 \pm 1.1$  & $6.4$\tablenotemark{e}\\
              & 2454934.46 & $102 \pm 9$ & $2.7 \pm 0.4$  & $7.5 \pm 1.1$ & $5.8 \pm 0.6$  & $6.4$\tablenotemark{e}\\
              & 2452938.41 & $106 \pm 9$ & \nodata        & \nodata       & $6.3 \pm 1.1$  & $6.4$\tablenotemark{e}\\
              & 2454938.53 & $106 \pm 9$ & $2.4 \pm 0.3$  & $6.2 \pm 0.9$ & $5.6 \pm 0.6$  & $6.3$\tablenotemark{e}\\
              & 2454944.38 & $112 \pm 9$ & $2.1 \pm 0.3$  & $5.8 \pm 1.2$ & $5.8 \pm 1.4$  & $5.6$\tablenotemark{e}\\
              & 2454946.40 & $114 \pm 9$ & \nodata        & \nodata       & $5.1 \pm 1.0$  & $5.0$\tablenotemark{e}\\
              & 2454955.45 & $123 \pm 9$ & $1.04 \pm 0.15$& $3.2 \pm 0.5$ & $3.5 \pm 0.9$  & $3.0$\tablenotemark{e}\\
              & 2454955.53 & $123 \pm 9$ & $1.07 \pm 0.15$& $3.3 \pm 0.5$ & $3.4 \pm 0.5$  & $2.9$\tablenotemark{e}\\
              & 2454957.42 & $125 \pm 9$ & \nodata        & \nodata       & \nodata        & $2.6$\tablenotemark{e}\\
              & 2454964.66 & $132 \pm 9$ & $0.72 \pm 0.10$& $2.4 \pm 0.4$ & $2.6 \pm 0.8$  & $2.0$\tablenotemark{e}\\
              & 2454971.52 & $139 \pm 9$ & $0.63 \pm 0.09$& $2.2 \pm 0.3$ & $2.8 \pm 0.8$  & $1.8$\tablenotemark{e}\\ 
              & 2454979.37 & $147 \pm 9$ & \nodata        & \nodata       & \nodata        & $1.7$\tablenotemark{e}\\
              & 2454982.59 & $150 \pm 9$ & $0.59 \pm 0.09$& $2.0 \pm 0.4$ & $2.1 \pm 1.4$  & $1.7$\tablenotemark{e}\\
              & 2455009.38 & $177 \pm 9$ & \nodata        & \nodata       & \nodata        & $1.2$\tablenotemark{e}\\
              & 2455031.39 & $199 \pm 9$ & \nodata        & \nodata       & \nodata        & $0.97$\tablenotemark{e}\\
              & 2455042.36 & $210 \pm 9$ & \nodata        & \nodata       & \nodata        & $0.90$\tablenotemark{e}\\
              & 2455056.35 & $224 \pm 9$ & $0.29 \pm 0.04$& $1.0 \pm 0.2$ & $0.6 \pm 0.2$  & $0.77$\tablenotemark{e}\\
              & 2455063.37 & $231 \pm 9$ & \nodata        & \nodata       & \nodata        & $0.72$\tablenotemark{e}\\
              & 2455072.34 & $240 \pm 9$ & $0.25 \pm 0.04$& $0.8 \pm 0.2$ & $0.49 \pm 0.18$& $0.64$\tablenotemark{e}\\
              & 2455079.34 & $247 \pm 9$ & $0.24 \pm 0.04$& $0.8 \pm 0.2$ & $0.48 \pm 0.14$& $0.63$\tablenotemark{e}\\
              & 2455154.69 & $322 \pm 9$ & \nodata        & \nodata       & \nodata        & $0.34$\tablenotemark{e}\\
              & 2455285.43 & $453 \pm 9$ & $0.042 \pm 0.008$ & $0.09 \pm 0.03$ & $0.079 \pm 0.019$ & \nodata\tablenotemark{e}\\
    \enddata 
    \tablerefs{\textsuperscript{a}\cite{pastorello_sn_2005}, \textsuperscript{b}\cite{hamuy_type_2001} \textsuperscript{c}\cite{kleiser_peculiar_2011}, \textsuperscript{d}\cite{taddia_type_2012}, \textsuperscript{e}\cite{pastorello_sn_2012}}
    \tablecomments{Reported uncertainties come from \texttt{SuperBoL}. Two significant figures are reported for uncertainties with a leading digit of 1. Uncertainties in literature values come from tables or plots in the referenced sources. Note that the reported photometry of SN~2000cb in \citet{kleiser_peculiar_2011} did not include the needed 4 photometric filters for \texttt{SuperBoL} to calculate the quasi-bolometric and direct integration luminosities.}
\end{deluxetable}
\capstarttrue

\clearpage

\capstartfalse
\begin{deluxetable}{cccccc}
    \tabletypesize{\footnotesize}
    \tablecolumns{5}
    \tablewidth{0pt}
    \tablecaption{Ejected $^{56}$Ni masses \label{tab:ni_mass}}
    \tablehead{
    \colhead{SN} & \colhead{qbol} & \colhead{Direct} & \colhead{BC} & \colhead{lit} & \colhead{Method}\\
                 & \colhead{(M$_{\odot}$)} & \colhead{(M$_{\odot}$)} & \colhead{(M$_{\odot}$)} & \colhead{(M$_{\odot}$)} &}
    \startdata
    1998A & $0.046 \pm 0.012$ & $0.15 \pm 0.04$ & $0.11 \pm 0.02$ &  0.11\tablenotemark{a} & D, SA\\
    2000cb & $0.032 \pm 0.015$ & $0.11 \pm 0.05$ &  $0.060 \pm 0.015$ & $0.09\pm0.037$\tablenotemark{b} & P\\
              &               &              &      & $0.1\pm0.02$\tablenotemark{c} & D, H\\
              &               &              &      & $0.083\pm0.039$\tablenotemark{d}& D, H\\
    2006V & $0.050 \pm 0.005$ & $0.17 \pm 0.017$ & $0.092 \pm 0.016$ & $0.127\pm0.010$\tablenotemark{e} & D\\
             &                 &               & &  $0.127$\tablenotemark{e} & D, SA\\
    2006au & --- & --- & --- & $\leq 0.073$\tablenotemark{e} & D, SA\\
    2009E & $0.017 \pm 0.001$ & $0.06 \pm 0.004$ & $0.048 \pm 0.005$ & 0.043\tablenotemark{f} & D, H\\
             &                 &               &  & 0.039\tablenotemark{f} & D, SA
    \enddata
    \tablerefs{\textsuperscript{a}\cite{pastorello_sn_2005}, \textsuperscript{b}\cite{hamuy_type_2001} \textsuperscript{c}\cite{kleiser_peculiar_2011}, \textsuperscript{d}\cite{utrobin_supernova_2011}, \textsuperscript{e}\cite{taddia_type_2012}, \textsuperscript{f}\cite{pastorello_sn_2012}}
    \tablecomments{Reported uncertainties are one standard deviation errors from the least squares fitting. Two significant figures are reported for uncertainties with a leading digit of 1. For each value of the $^{56}$Ni mass from the literature, the method used to calculate it has also been reported. D: $L_{\text{bol}}$ from direct integration. P: $L_{\text{bol}}$ from polynomial fits. H: M($^{56}$Ni) from hydrodynamic models. SA: M($^{56}$Ni) from semi-analytic models.}
\end{deluxetable}
\capstarttrue

\capstartfalse
\begin{deluxetable}{ccccc}
    \tabletypesize{\footnotesize}
    \tablecolumns{6}
    \tablewidth{0pt}
    \tablecaption{Peak observed SN luminosities \label{tab:sn_Lpeak}}
    \tablehead{\colhead{SN} & \colhead{$L_{\text{qbol}}$} & \colhead{$L_{\text{D}}$} & \colhead{$L_\text{BC}$} & \colhead{$L_{\text{lit}}$} \\
                 & $10^{41}$ (erg/s) & $10^{41}$ (erg/s) & $10^{41}$ (erg/s) & $10^{41}$ (erg/s)}
    \startdata
    1998A & $5\pm2$ & $13\pm6$ & $12\pm3$ & $6.3\pm0.8$\tablenotemark{a} \\
    2000cb & $4\pm1.8$ & $12\pm6$ & $12\pm4$ & $9.4$\tablenotemark{b} \\
    2006V & $8.5\pm1.2$ & $22\pm3$ & $18\pm3$ & $18.2\pm0.4$\tablenotemark{c}\\
    2006au & $7.7\pm1.1$ & $20\pm3$ & $15\pm3$ & $16.2\pm0.6$\tablenotemark{c}\\
    2009E & $2.7\pm0.4$ & $7.5\pm1.1$ & $6.5\pm1.2$ & $6.4$\tablenotemark{d}
    \enddata
    \tablerefs{\textsuperscript{a}\cite{pastorello_sn_2005}, \textsuperscript{b}\cite{hamuy_type_2001}, \textsuperscript{c}\cite{taddia_type_2012}, \textsuperscript{d}\cite{pastorello_sn_2012}}
\end{deluxetable}
\capstarttrue

\clearpage

\capstartfalse
\begin{deluxetable}{cccc}
    \tabletypesize{\footnotesize}
    \tablecolumns{4}
    \tablewidth{0pt}
    \tablecaption{Bolometric Luminosity with Sparse Photometry\label{tab:sparse_phot}}
    \tablehead{\colhead{Bands} & \colhead{$L_\text{qbol}$} & \colhead{$L_\text{D}$} & \colhead{$L_\text{BC}$}\\
                    & $10^{41}$ (erg/s) & $10^{41}$ (erg/s) & $10^{41}$ (erg/s)}
    \startdata
    VRI  & --- & ---  & $6.18$ \\
    BVR  & --- & ---  & $12.52$ \\
    UBV  & --- & ---  & $12.52$ \\
    BVRI & $3.49$ & $10.54$ & $8.62$ \\
    UBVR & $3.78$ & $10.51$ & $12.52$ \\
    UBVRI & $4.88$ & $10.64$ & $8.62$ \\
    UBVRIJHK & $7.25$ & $11.59$ & $8.62$ \\
    \hline
    \\
    $L_{\text{bol}}$ (\texttt{PHOENIX}) & & $11.08$
    \enddata
\end{deluxetable}
\capstarttrue

\end{document}